\documentclass[10pt, conference, letterpaper, nofonttune]{IEEEtran}

\usepackage[utf8]{inputenc}
\usepackage[english]{babel}
\usepackage{diagbox}
\usepackage[numbers,sort&compress]{natbib}

\usepackage{url}  

\usepackage{epsfig}
\usepackage{epstopdf}
\usepackage{tikz}
\usepackage{lscape}
\usepackage{amsmath}
\usepackage{amssymb}
\usepackage{amsfonts}
\usepackage{amsthm}
\usepackage{fancyvrb}
\usepackage{mathtools}
\usepackage{verbatim}
\usepackage{xfrac}
\usepackage{algorithmic}
\usepackage{amsmath}
\usepackage{multirow}
\usepackage{algorithm}
\usepackage{proof}
\usepackage{graphics}
\usepackage{graphicx}
\usepackage{subfigure}
\usetikzlibrary{arrows,automata}
\usetikzlibrary{shapes}
\usepackage{enumitem}
\usepackage{xspace}
\usepackage{pifont}
\usepackage{soul}
\usepackage{balance}

\newcommand{\mods}{\mathcal{M}}
\newcommand{\funcs}{\mathcal{F}}
\newcommand{\devs}{\mathcal{D}}
\newcommand{\reqs}{\mathcal{I}}
\newcommand{\types}{\mathcal{T}}
\newcommand{\tin}{t^{\mathrm{IN}}}

\newcommand{\oran}{O-RAN\xspace}
\newcommand{\name}{OrchestRAN\xspace}
\newcommand{\nonrt}{non-\gls{rt}\xspace}
\newcommand{\nearrt}{near-\gls{rt}\xspace}
\newcommand{\rt}{\gls{rt}\xspace}

\newcommand{\cmark}{\ding{51}}%
\newcommand{\xmark}{\ding{55}}%

\newif\ifCompile
\Compiletrue

\usepackage[acronyms,nonumberlist,nopostdot,nomain,nogroupskip,acronymlists={hidden}]{glossaries}
\newglossary[algh]{hidden}{acrh}{acnh}{Hidden Acronyms}
\newacronym{3gpp}{3GPP}{3rd Generation Partnership Project}
\newacronym{4g}{4G}{4th generation}
\newacronym{5g}{5G}{5th generation}
\newacronym{5gc}{5GC}{5G Core}
\newacronym{adc}{ADC}{Analog to Digital Converter}
\newacronym{aerpaw}{AERPAW}{Aerial Experimentation and Research Platform for Advanced Wireless}
\newacronym{ai}{AI}{Artificial Intelligence}
\newacronym{aimd}{AIMD}{Additive Increase Multiplicative Decrease}
\newacronym{am}{AM}{Acknowledged Mode}
\newacronym{amc}{AMC}{Adaptive Modulation and Coding}
\newacronym{amf}{AMF}{Access and Mobility Management Function}
\newacronym{aops}{AOPS}{Adaptive Order Prediction Scheduling}
\newacronym{api}{API}{Application Programming Interface}
\newacronym{apn}{APN}{Access Point Name}
\newacronym{ap}{AP}{application protocol}
\newacronym{aqm}{AQM}{Active Queue Management}
\newacronym{ausf}{AUSF}{Authentication Server Function}
\newacronym{avc}{AVC}{Advanced Video Coding}
\newacronym{awgn}{AGWN}{Additive White Gaussian Noise}
\newacronym{balia}{BALIA}{Balanced Link Adaptation Algorithm}
\newacronym{bbu}{BBU}{Base Band Unit}
\newacronym{bdp}{BDP}{Bandwidth-Delay Product}
\newacronym{ber}{BER}{Bit Error Rate}
\newacronym{bf}{BF}{Beamforming}
\newacronym{bler}{BLER}{Block Error Rate}
\newacronym{brr}{BRR}{Bayesian Ridge Regressor}
\newacronym{bs}{BS}{Base Station}
\newacronym{bsr}{BSR}{Buffer Status Report}
\newacronym{bss}{BSS}{Business Support System}
\newacronym{ca}{CA}{Carrier Aggregation}
\newacronym{caas}{CaaS}{Connectivity-as-a-Service}
\newacronym{cb}{CB}{Code Block}
\newacronym{cc}{CC}{Congestion Control}
\newacronym{ccid}{CCID}{Congestion Control ID}
\newacronym{cco}{CC}{Carrier Component}
\newacronym{cdd}{CDD}{Cyclic Delay Diversity}
\newacronym{cdf}{CDF}{Cumulative Distribution Function}
\newacronym{cdn}{CDN}{Content Distribution Network}
\newacronym{cn}{CN}{Core Network}
\newacronym{codel}{CoDel}{Controlled Delay Management}
\newacronym{comac}{COMAC}{Converged Multi-Access and Core}
\newacronym{cord}{CORD}{Central Office Re-architected as a Datacenter}
\newacronym{cornet}{CORNET}{COgnitive Radio NETwork}
\newacronym{cosmos}{COSMOS}{Cloud Enhanced Open Software Defined Mobile Wireless Testbed for City-Scale Deployment}
\newacronym{cots}{COTS}{Commercial Off-the-Shelf}
\newacronym{cp}{CP}{Control Plane}
\newacronym{cpu}{CPU}{Central Processing Unit}
\newacronym{cqi}{CQI}{Channel Quality Information}
\newacronym{cr}{CR}{Cognitive Radio}
\newacronym{cran}{CRAN}{Cloud \gls{ran}}
\newacronym{crs}{CRS}{Cell Reference Signal}
\newacronym{csi}{CSI}{Channel State Information}
\newacronym{csirs}{CSI-RS}{Channel State Information - Reference Signal}
\newacronym{cu}{CU}{Central Unit}
\newacronym{d2tcp}{D$^2$TCP}{Deadline-aware Data center TCP}
\newacronym{d3}{D$^3$}{Deadline-Driven Delivery}
\newacronym{dac}{DAC}{Digital to Analog Converter}
\newacronym{dag}{DAG}{Directed Acyclic Graph}
\newacronym{das}{DAS}{Distributed Antenna System}
\newacronym{dash}{DASH}{Dynamic Adaptive Streaming over HTTP}
\newacronym{dc}{DC}{Dual Connectivity}
\newacronym{dccp}{DCCP}{Datagram Congestion Control Protocol}
\newacronym{dce}{DCE}{Direct Code Execution}
\newacronym{dci}{DCI}{Downlink Control Information}
\newacronym{dctcp}{DCTCP}{Data Center TCP}
\newacronym{dl}{DL}{Downlink}
\newacronym{dmr}{DMR}{Deadline Miss Ratio}
\newacronym{dmrs}{DMRS}{DeModulation Reference Signal}
\newacronym{dnn}{DNN}{Deep Neural Network}
\newacronym{drlcc}{DRL-CC}{Deep Reinforcement Learning Congestion Control}
\newacronym{drs}{DRS}{Discovery Reference Signal}
\newacronym{du}{DU}{Distributed Unit}
\newacronym{e2e}{E2E}{end-to-end}
\newacronym{ecaas}{ECaaS}{Edge-Cloud-as-a-Service}
\newacronym{ecn}{ECN}{Explicit Congestion Notification}
\newacronym{edf}{EDF}{Earliest Deadline First}
\newacronym{embb}{eMBB}{Enhanced Mobile Broadband}
\newacronym{empower}{EMPOWER}{EMpowering transatlantic PlatfOrms for advanced WirEless Research}
\newacronym{enb}{eNB}{evolved Node Base}
\newacronym{endc}{EN-DC}{E-UTRAN-\gls{nr} \gls{dc}}
\newacronym{epc}{EPC}{Evolved Packet Core}
\newacronym{eps}{EPS}{Evolved Packet System}
\newacronym{es}{ES}{Edge Server}
\newacronym{etsi}{ETSI}{European Telecommunications Standards Institute}
\newacronym[firstplural=Estimated Times of Arrival (ETAs)]{eta}{ETA}{Estimated Time of Arrival}
\newacronym{eutran}{E-UTRAN}{Evolved Universal Terrestrial Access Network}
\newacronym{faas}{FaaS}{Function-as-a-Service}
\newacronym{fapi}{FAPI}{Functional Application Platform Interface}
\newacronym{fdd}{FDD}{Frequency Division Duplexing}
\newacronym{fdm}{FDM}{Frequency Division Multiplexing}
\newacronym{fdma}{FDMA}{Frequency Division Multiple Access}
\newacronym{fed4fire}{FED4FIRE+}{Federation 4 Future Internet Research and Experimentation Plus}
\newacronym{fir}{FIR}{finite impulse response}
\newacronym{fit}{FIT}{Future \acrlong{iot}}
\newacronym{fpga}{FPGA}{Field Programmable Gate Array}
\newacronym{fr2}{FR2}{Frequency Range 2}
\newacronym{fs}{FS}{Fast Switching}
\newacronym{fscc}{FSCC}{Flow Sharing Congestion Control}
\newacronym{ftp}{FTP}{File Transfer Protocol}
\newacronym{fw}{FW}{Flow Window}
\newacronym{ge}{GE}{Gaussian Elimination}
\newacronym{gnb}{gNB}{Next Generation Node Base}
\newacronym{gop}{GOP}{Group of Pictures}
\newacronym{gpr}{GPR}{Gaussian Process Regressor}
\newacronym{gpu}{GPU}{Graphics Processing Unit}
\newacronym{gtp}{GTP}{GPRS Tunneling Protocol}
\newacronym{gtpc}{GTP-C}{GPRS Tunnelling Protocol Control Plane}
\newacronym{gtpu}{GTP-U}{GPRS Tunnelling Protocol User Plane}
\newacronym{gtpv2c}{GTPv2-C}{\gls{gtp} v2 - Control}
\newacronym{gw}{GW}{Gateway}
\newacronym{harq}{HARQ}{Hybrid Automatic Repeat reQuest}
\newacronym{hetnet}{HetNet}{Heterogeneous Network}
\newacronym{hh}{HH}{Hard Handover}
\newacronym{hol}{HOL}{Head-of-Line}
\newacronym{hqf}{HQF}{Highest-quality-first}
\newacronym{hss}{HSS}{Home Subscription Server}
\newacronym{http}{HTTP}{HyperText Transfer Protocol}
\newacronym{ia}{IA}{Initial Access}
\newacronym{iab}{IAB}{Integrated Access and Backhaul}
\newacronym{ic}{IC}{Incident Command}
\newacronym{ietf}{IETF}{Internet Engineering Task Force}
\newacronym{imsi}{IMSI}{International Mobile Subscriber Identity}
\newacronym{imt}{IMT}{International Mobile Telecommunication}
\newacronym{idn}{IDN}{Inference Delivery Networks}
\newacronym{iot}{IoT}{Internet of Things}
\newacronym{ip}{IP}{Internet Protocol}
\newacronym{itu}{ITU}{International Telecommunication Union}
\newacronym{kpi}{KPI}{Key Performance Indicator}
\newacronym{kpm}{KPM}{Key Performance Measurement}
\newacronym{kvm}{KVM}{Kernel-based Virtual Machine}
\newacronym{los}{LOS}{Line-of-Sight}
\newacronym{lsm}{LSM}{Link-to-System Mapping}
\newacronym{lstm}{LSTM}{Long Short Term Memory}
\newacronym{lte}{LTE}{Long Term Evolution}
\newacronym{lxc}{LXC}{Linux Container}
\newacronym{m2m}{M2M}{Machine to Machine}
\newacronym{mac}{MAC}{Medium Access Control}
\newacronym{manet}{MANET}{Mobile Ad Hoc Network}
\newacronym{mano}{MANO}{management~and orchestration}
\newacronym{mc}{MC}{Multi-Connectivity}
\newacronym{mcc}{MCC}{Mobile Cloud Computing}
\newacronym{mchem}{MCHEM}{Massive Channel Emulator}
\newacronym{mcs}{MCS}{Modulation and Coding Scheme}
\newacronym{mec}{MEC}{Multi-access Edge Computing}
\newacronym{mec2}{MEC}{Mobile Edge Cloud}
\newacronym{mfc}{MFC}{Mobile Fog Computing}
\newacronym{mi}{MI}{Mutual Information}
\newacronym{mib}{MIB}{Master Information Block}
\newacronym{miesm}{MIESM}{Mutual Information Based Effective SINR}
\newacronym{mimo}{MIMO}{Multiple Input, Multiple Output}
\newacronym{ml}{ML}{Machine Learning}
\newacronym{mlr}{MLR}{Maximum-local-rate}
\newacronym[plural=\gls{mme}s,firstplural=Mobility Management Entities (MMEs)]{mme}{MME}{Mobility Management Entity}
\newacronym{mmtc}{mMTC}{Massive Machine-Type Communications}
\newacronym{mmwave}{mmWave}{millimeter wave}
\newacronym{mpdccp}{MP-DCCP}{Multipath Datagram Congestion Control Protocol}
\newacronym{mptcp}{MPTCP}{Multipath TCP}
\newacronym{mr}{MR}{Maximum Rate}
\newacronym{mrdc}{MR-DC}{Multi \gls{rat} \gls{dc}}
\newacronym{mse}{MSE}{Mean Square Error}
\newacronym{mss}{MSS}{Maximum Segment Size}
\newacronym{mt}{MT}{Mobile Termination}
\newacronym{mtd}{MTD}{Machine-Type Device}
\newacronym{mtu}{MTU}{Maximum Transmission Unit}
\newacronym{mumimo}{MU-MIMO}{Multi-user \gls{mimo}}
\newacronym{mvno}{MVNO}{Mobile Virtual Network Operator}
\newacronym{nalu}{NALU}{Network Abstraction Layer Unit}
\newacronym{nas}{NAS}{Non-Access Stratum}
\newacronym{nbiot}{NB-IoT}{Narrow Band IoT}
\newacronym{nfv}{NFV}{Network Function Virtualization}
\newacronym{nfvi}{NFVI}{Network Function Virtualization Infrastructure}
\newacronym{nic}{NIC}{Network Interface Card}
\newacronym{nlos}{NLOS}{Non-Line-of-Sight}
\newacronym{no}{NO}{Network Operator}
\newacronym{now}{NOW}{Non Overlapping Window}
\newacronym{nsm}{NSM}{Network Service Mesh}
\newacronym[type=hidden]{nr}{NR}{New Radio}
\newacronym{nrf}{NRF}{Network Repository Function}
\newacronym{nsa}{NSA}{Non Stand Alone}
\newacronym{nse}{NSE}{Network Slicing Engine}
\newacronym{nssf}{NSSF}{Network Slice Selection Function}
\newacronym{o2i}{O2I}{Outdoor to Indoor}
\newacronym{oai}{OAI}{OpenAirInterface}
\newacronym{oaicn}{OAI-CN}{\gls{oai} \acrlong{cn}}
\newacronym{oairan}{OAI-RAN}{\acrlong{oai} \acrlong{ran}}
\newacronym{oam}{OAM}{Operations, Administration and Maintenance}
\newacronym{ofdm}{OFDM}{Orthogonal Frequency Division Multiplexing}
\newacronym{olia}{OLIA}{Opportunistic Linked Increase Algorithm}
\newacronym{omec}{OMEC}{Open Mobile Evolved Core}
\newacronym{onap}{ONAP}{Open Network Automation Platform}
\newacronym{onf}{ONF}{Open Networking Foundation}
\newacronym{onos}{ONOS}{Open Networking Operating System}
\newacronym{oom}{OOM}{\gls{onap} Operations Manager}
\newacronym{opnfv}{OPNFV}{Open Platform for \gls{nfv}}
\newacronym[type=hidden]{oran}{O-RAN}{Open \gls{ran}}
\newacronym{orbit}{ORBIT}{Open-Access Research Testbed for Next-Generation Wireless Networks}
\newacronym{os}{OS}{Operating System}
\newacronym{oss}{OSS}{Operations Support System}
\newacronym{pa}{PA}{Position-aware}
\newacronym{pase}{PASE}{Prioritization, Arbitration, and Self-adjusting Endpoints}
\newacronym{pawr}{PAWR}{Platforms for Advanced Wireless Research}
\newacronym{pbch}{PBCH}{Physical Broadcast Channel}
\newacronym{pcef}{PCEF}{Policy and Charging Enforcement Function}
\newacronym{pcfich}{PCFICH}{Physical Control Format Indicator Channel}
\newacronym{pcrf}{PCRF}{Policy and Charging Rules Function}
\newacronym{pdcch}{PDCCH}{Physical Downlink Control Channel}
\newacronym{pdcp}{PDCP}{Packet Data Convergence Protocol}
\newacronym{pdsch}{PDSCH}{Physical Downlink Shared Channel}
\newacronym{pdu}{PDU}{Packet Data Unit}
\newacronym{pf}{PF}{Proportional Fair}
\newacronym{pgw}{PGW}{Packet Gateway}
\newacronym{phich}{PHICH}{Physical Hybrid ARQ Indicator Channel}
\newacronym{phy}{PHY}{Physical}
\newacronym{pmch}{PMCH}{Physical Multicast Channel}
\newacronym{pmi}{PMI}{Precoding Matrix Indicators}
\newacronym{powder}{POWDER}{Platform for Open Wireless Data-driven Experimental Research}
\newacronym{ppo}{PPO}{Proximal Policy Optimization}
\newacronym{ppp}{PPP}{Poisson Point Process}
\newacronym{prach}{PRACH}{Physical Random Access Channel}
\newacronym{prb}{PRB}{Physical Resource Block}
\newacronym{psnr}{PSNR}{Peak Signal to Noise Ratio}
\newacronym{pss}{PSS}{Primary Synchronization Signal}
\newacronym{pucch}{PUCCH}{Physical Uplink Control Channel}
\newacronym{pusch}{PUSCH}{Physical Uplink Shared Channel}
\newacronym{qam}{QAM}{Quadrature Amplitude Modulation}
\newacronym{qci}{QCI}{\gls{qos} Class Identifier}
\newacronym{qoe}{QoE}{Quality of Experience}
\newacronym{qos}{QoS}{Quality of Service}
\newacronym{quic}{QUIC}{Quick UDP Internet Connections}
\newacronym{rach}{RACH}{Random Access Channel}
\newacronym{ran}{RAN}{Radio Access Network}
\newacronym[firstplural=Radio Access Technologies (RATs)]{rat}{RAT}{Radio Access Technology}
\newacronym{rcn}{RCN}{Research Coordination Network}
\newacronym{rec}{REC}{Radio Edge Cloud}
\newacronym{red}{RED}{Random Early Detection}
\newacronym{renew}{RENEW}{Reconfigurable Eco-system for Next-generation End-to-end Wireless}
\newacronym{rf}{RF}{Radio Frequency}
\newacronym{rfc}{RFC}{Request for Comments}
\newacronym{rfr}{RFR}{Random Forest Regressor}
\newacronym{ric}{RIC}{\gls{ran} Intelligent Controller}
\newacronym{rlc}{RLC}{Radio Link Control}
\newacronym{rlf}{RLF}{Radio Link Failure}
\newacronym{rlnc}{RLNC}{Random Linear Network Coding}
\newacronym{rmr}{RMR}{RIC Message Router}
\newacronym{rmse}{RMSE}{Root Mean Squared Error}
\newacronym{rnis}{RNIS}{Radio Network Information Service}
\newacronym{rr}{RR}{Round Robin}
\newacronym{rrc}{RRC}{Radio Resource Control}
\newacronym{rrm}{RRM}{Radio Resource Management}
\newacronym{rru}{RRU}{Remote Radio Unit}
\newacronym{rs}{RS}{Remote Server}
\newacronym{rsrp}{RSRP}{Reference Signal Received Power}
\newacronym{rsrq}{RSRQ}{Reference Signal Received Quality}
\newacronym{rss}{RSS}{Received Signal Strength}
\newacronym{rssi}{RSSI}{Received Signal Strength Indicator}
\newacronym{rt}{RT}{Real-time}
\newacronym{rtt}{RTT}{Round Trip Time}
\newacronym{ru}{RU}{Radio Unit}
\newacronym{rw}{RW}{Receive Window}
\newacronym{rx}{RX}{Receiver}
\newacronym{s1ap}{S1AP}{S1 Application Protocol}
\newacronym{sa}{SA}{standalone}
\newacronym{sack}{SACK}{Selective Acknowledgment}
\newacronym{sap}{SAP}{Service Access Point}
\newacronym{sc2}{SC2}{Spectrum Collaboration Challenge}
\newacronym{scef}{SCEF}{Service Capability Exposure Function}
\newacronym{sch}{SCH}{Secondary Cell Handover}
\newacronym{scoot}{SCOOT}{Split Cycle Offset Optimization Technique}
\newacronym{sctp}{SCTP}{Stream Control Transmission Protocol}
\newacronym{sdap}{SDAP}{Service Data Adaptation Protocol}
\newacronym{sdk}{SDK}{Software Development Kit}
\newacronym{sdm}{SDM}{Space Division Multiplexing}
\newacronym{sdma}{SDMA}{Spatial Division Multiple Access}
\newacronym{sdn}{SDN}{Software-defined Networking}
\newacronym{sdr}{SDR}{Software-defined Radio}
\newacronym{seba}{SEBA}{SDN-Enabled Broadband Access}
\newacronym{sgsn}{SGSN}{Serving GPRS Support Node}
\newacronym{sgw}{SGW}{Service Gateway}
\newacronym{si}{SI}{Study Item}
\newacronym{sib}{SIB}{Secondary Information Block}
\newacronym{sinr}{SINR}{Signal to Interference plus Noise Ratio}
\newacronym{sip}{SIP}{Session Initiation Protocol}
\newacronym{siso}{SISO}{Single Input, Single Output}
\newacronym{sla}{SLA}{service level agreement}
\newacronym{sm}{SM}{Service Model}
\newacronym{smf}{SMF}{Session Management Function}
\newacronym{smo}{SMO}{service management and orchestration}
\newacronym{sms}{SMS}{Short Message Service}
\newacronym{smsgmsc}{SMS-GMSC}{\gls{sms}-Gateway}
\newacronym{snr}{SNR}{Signal-to-Noise-Ratio}
\newacronym{son}{SON}{Self-Organizing Network}
\newacronym{sptcp}{SPTCP}{Single Path TCP}
\newacronym{srb}{SRB}{Service Radio Bearer}
\newacronym{srn}{SRN}{Standard Radio Node}
\newacronym{srs}{SRS}{Sounding Reference Signal}
\newacronym{ss}{SS}{Synchronization Signal}
\newacronym{sss}{SSS}{Secondary Synchronization Signal}
\newacronym{st}{ST}{Spanning Tree}
\newacronym{svc}{SVC}{Scalable Video Coding}
\newacronym{tb}{TB}{Transport Block}
\newacronym{tcp}{TCP}{Transmission Control Protocol}
\newacronym{tdd}{TDD}{Time Division Duplexing}
\newacronym{tdm}{TDM}{Time Division Multiplexing}
\newacronym{tdma}{TDMA}{Time Division Multiple Access}
\newacronym{tfl}{TfL}{Transport for London}
\newacronym{tfrc}{TFRC}{TCP-Friendly Rate Control}
\newacronym{tft}{TFT}{Traffic Flow Template}
\newacronym{tgen}{TGEN}{Traffic Generator}
\newacronym{tip}{TIP}{Telecom Infra Project}
\newacronym{tm}{TM}{Transparent Mode}
\newacronym[plural=Telcos,firstplural=Telecommunications Companies (Telcos)]{to}{Telco}{Telecommunications Company}
\newacronym{tr}{TR}{Technical Report}
\newacronym{trp}{TRP}{Transmitter Receiver Pair}
\newacronym{ts}{TS}{Technical Specification}
\newacronym{tti}{TTI}{Transmission Time Interval}
\newacronym{ttt}{TTT}{Time-to-Trigger}
\newacronym{tx}{TX}{Transmitter}
\newacronym{uas}{UAS}{Unmanned Aerial System}
\newacronym{uav}{UAV}{Unmanned Aerial Vehicle}
\newacronym{udm}{UDM}{Unified Data Management}
\newacronym{udp}{UDP}{User Datagram Protocol}
\newacronym{udr}{UDR}{Unified Data Repository}
\newacronym{ue}{UE}{User Equipment}
\newacronym{uhd}{UHD}{\gls{usrp} Hardware Driver}
\newacronym{ul}{UL}{Uplink}
\newacronym{um}{UM}{Unacknowledged Mode}
\newacronym{uml}{UML}{Unified Modeling Language}
\newacronym{upa}{UPA}{Uniform Planar Array}
\newacronym{upf}{UPF}{User Plane Function}
\newacronym{urllc}{URLLC}{Ultra Reliable and Low Latency Communication}
\newacronym{usa}{U.S.}{United States}
\newacronym{usim}{USIM}{Universal Subscriber Identity Module}
\newacronym{usrp}{USRP}{Universal Software Radio Peripheral}
\newacronym{utc}{UTC}{Urban Traffic Control}
\newacronym{vim}{VIM}{Virtualization Infrastructure Manager}
\newacronym{vm}{VM}{Virtual Machine}
\newacronym{vnf}{VNF}{Virtual Network Function}
\newacronym{volte}{VoLTE}{Voice over \gls{lte}}
\newacronym{voltha}{VOLTHA}{Virtual OLT HArdware Abstraction}
\newacronym{vr}{VR}{Virtual Reality}
\newacronym{vran}{vRAN}{Virtualized \gls{ran}}
\newacronym{vss}{VSS}{Video Streaming Server}
\newacronym{wbf}{WBF}{Wired Bias Function}
\newacronym{wf}{WF}{Wired-first}
\newacronym{wlan}{WLAN}{Wireless Local Area Network}
\newacronym{osm}{OSM}{Open Source \gls{nfv} Management and Orchestration}
\newacronym{pnf}{PNF}{Physical Network Function}
\newacronym{drl}{DRL}{Deep Reinforcement Learning}
\newacronym{mtc}{MTC}{Machine-type Communications}

\glsdisablehyper

\title{OrchestRAN: Network Automation through Orchestrated Intelligence in the Open RAN}

\author{\IEEEauthorblockN{
Salvatore D'Oro,
Leonardo Bonati,
Michele Polese,
and Tommaso Melodia}
\IEEEauthorblockA{Institute for the Wireless Internet of Things, Northeastern University, Boston, MA, U.S.A.\vspace{-1.5cm}}
\thanks{Email: \{s.doro, l.bonati, m.polese, melodia\}@northeastern.edu}
\thanks{This work was partially supported by the U.S.\ National Science Foundation under Grants CNS-1923789 and NSF CNS-1925601, and the U.S.\ Office of Naval Research under Grant N00014-20-1-2132.}
}

\IEEEoverridecommandlockouts

\newlist{enumeratebyten}{enumerate}{1}
\setlist[enumeratebyten]{label={\textbf{Q\arabic*:}},ref={Q\arabic*:}}

\begin{document}

\maketitle

\glsunset{mac}

\begin{abstract}

The next generation of cellular networks will be characterized by softwarized, open, and disaggregated architectures exposing analytics and control knobs to
enable network intelligence via innovative data-driven algorithms.
%
How to practically realize this vision, however, is largely an open problem. For a given network optimization/automation objective, it is currently unknown how to select which data-driven models should be deployed and where,
which parameters to control, and how to feed them appropriate inputs. 
In this paper, we take a decisive step forward by presenting and prototyping \name, a novel orchestration framework for next generation systems that embraces and builds upon the Open \gls{ran} paradigm
to provide a practical solution to these challenges.
\name has been designed to execute in the \nonrt \gls{ric} and allows Network Operators (NOs) to specify high-level control/inference objectives (i.e., adapt scheduling, and forecast capacity in \nearrt, e.g.,  for a set of base stations in Downtown New York). \name automatically computes the optimal set of data-driven algorithms and their execution location (e.g., in the cloud, or at the edge) to achieve intents specified by the NOs while meeting the desired timing requirements and avoiding conflicts between different data-driven algorithms controlling the same parameters set.
%
%
We show that the intelligence orchestration problem in Open RAN is NP-hard, and design low-complexity solutions to support real-world applications. We prototype \name and test it at scale on Colosseum, the world's largest wireless network emulator with hardware in the loop. Our experimental results on a network with 7 base stations and 42 users demonstrate that \name is able to instantiate data-driven services on demand with minimal control overhead and latency.
\end{abstract}


\begin{picture}(0,0)(10,-450)
\put(0,0){
\put(0,10){\footnotesize This paper has been accepted for publication on IEEE International Conference on Computer Communications (INFOCOM) 2022.}
\put(0,0){\tiny \copyright 2022 IEEE. Personal use of this material is permitted. Permission from IEEE must be obtained for all other uses, in any current or future media including reprinting/republishing}
\put(0,-5){\tiny this material for advertising or promotional purposes, creating new collective works, for resale or redistribution to servers or lists, or reuse of any copyrighted component of this work in other works.}
\put(0,-20){\scriptsize }}
\end{picture}

\glsresetall
\glsunset{mac}

\vspace{-0.7cm}
\section{Introduction} \label{sec:intro}

The fifth-generation (5G) of cellular networks and its evolution (NextG), will mark the end of the era of inflexible hardware-based \gls{ran} architectures in favor of innovative and agile solutions built upon softwarization, openness and disaggregation principles. This paradigm shift---often referred to as Open \gls{ran}---comes with unprecedented flexibility. It makes it possible to split network functionalities---traditionally embedded and executed in monolithic base stations---and instantiate and control them across multiple nodes of the network~\cite{bonati2020open}.

\begin{figure}[t]
\centering
    \includegraphics[width=0.9\columnwidth]{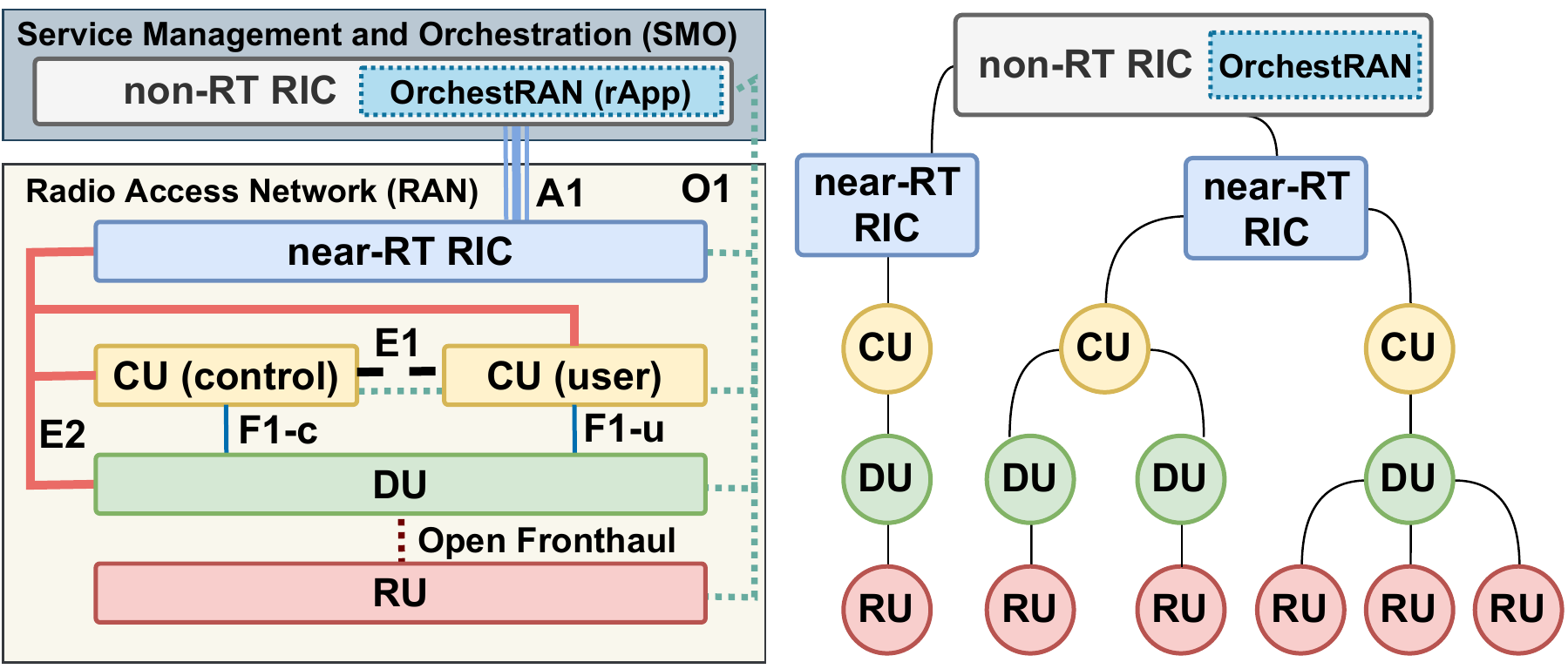}
    \vspace{-0.3cm}
    \caption{\oran reference architecture and interfaces (left). Representation of an \oran network architecture as a tree graph (right).}
    \label{fig:system_split}
\end{figure}

In this context, the \oran Alliance~\cite{oran-arch-spec}, a consortium led by \glspl{no}, vendors and academic partners, is developing a standardized architecture for Open RAN that promotes horizontal disaggregation and standardization of \gls{ran} interfaces, thus enabling multivendor equipment interoperability and  algorithmic network control and analytics.
As shown in Fig.~\ref{fig:system_split}, \oran embraces the \gls{3gpp} functional split with \glspl{cu}, \glspl{du} and \glspl{ru} implementing different functions of the protocol stack. \oran also introduces (i) a set of open standardized interfaces to interact, control and collect data from every node of the network; as well as (ii) \glspl{ric} that execute third-party applications over an abstract overlay to control \gls{ran} functionalities, i.e., \textit{xApps} in the \nearrt and \textit{rApps} in the \nonrt \gls{ric}. The \oran  architecture makes it possible to bring automation and intelligence to the network through \gls{ml} and \gls{ai},
which will leverage the enormous amount of data generated by the \gls{ran}---and exposed through the \oran interfaces---to analyze the current network conditions, forecast future traffic profiles and demand, and implement closed-loop network control strategies to optimize the \gls{ran} performance.
For this reason, how to design, train and deploy reliable and effective data-driven solutions has recently received increasing interest from academia and industry alike,
with applications ranging from controlling \gls{ran} resource and transmission policies~\cite{9206115, 9468707,okic2020pi,bega2020network,bonati2021intelligence,9490293,salhab2021autonomous,9459763,9286741,bonati2020cellos,casasole2021qcell}, to forecasting and classifying traffic and \glspl{kpi}~\cite{bega2019deepcog,paul2019traffic,polese2021deepbeam,klautau20185g,luo2018channel},
thus highlighting how these approaches will be foundational to the Open RAN paradigm.
However, how to deploy and manage, i.e., \textit{orchestrate}, intelligence into softwarized cellular networks is by no means a solved problem for the following reasons:

\noindent
$\bullet$ \textit{Complying with time scales and making input available:} Adapting \gls{ran} parameters and functionalities requires control loops operating over time scales ranging from a few milliseconds (i.e., real-time) to a few hundreds of milliseconds (i.e., \nearrt) to several seconds (i.e., \nonrt)~\cite{9447770,bonati2021intelligence}.
As a consequence, the models and the location where they are executed need
to be selected to be able to retrieve the necessary inputs and compute the output within the appropriate time constraints~\cite{bonati2021intelligence,salem2021towards}.
For instance, while IQ samples are easily available in real time at the \gls{ran}, it is extremely hard to make them available at the near-RT and non-RT \glspl{ric} within the same temporal window, making the execution of models that require IQ samples as input on the \glspl{ric} ineffective.

\noindent
$\bullet$ \textit{Choosing the right model:} Each ML/AI model is designed to accomplish specific inference and/or control tasks and requires well-defined inputs in terms of data type and size. One must make sure that the most suitable model is selected for a specific \gls{no} request, and that it meets the required performance metrics (e.g., minimum accuracy), delivers the desired inference/control functionalities, and is instantiated on nodes with enough resources to execute it.

\noindent
$\bullet$ \textit{Conflict mitigation:} One must also ensure that selected ML/AI models do not conflict with each other, and that the same parameter (or functionality) is controlled by only a single model at any given time.

For these reasons, \textit{orchestrating} network intelligence in the Open RAN presents unprecedented and unique challenges that call for innovative, automated and scalable solutions.
%
In this paper, we address these challenges by presenting \name, an automated intelligence orchestration framework for
the Open \gls{ran}.
%
\name follows the \oran specifications and operates as an rApp executed in the \nonrt \gls{ric} (Fig.~\ref{fig:system_split}) providing automated routines to: (i)~Collect control requests from \glspl{no}; (ii)~select the optimal ML/AI models to achieve \glspl{no}' goals and avoid conflicts; (iii)~determine the optimal execution location for each model complying with timescale requirements, resource and data availability,
and (iv)~automatically embed models into \oran applications that are dispatched to selected nodes, where they are executed and fed the required inputs.

To achieve this goal, we have designed and prototyped novel algorithms embedding pre-processing variable reduction and branching techniques that allow \name to compute orchestration solutions with different complexity and optimality trade-offs, while ensuring that the \glspl{no} intents are satisfied.
We evaluate the performance of \name in orchestrating intelligence in the \gls{ran} through numerical simulations, and by prototyping \name on ColO-RAN~\cite{polese2021coloran}, an O-RAN-compliant large-scale experimental platform developed on top of Colosseum, the world's largest wireless network emulator with hardware in-the-loop~\cite{bonati2021colosseum}. Experimental results on an \oran-compliant softwarized network with 7 cellular base stations and 42 users demonstrate that \name enables seamless instantiation of \oran applications with diverse timescale requirements at different \oran components. \name automatically selects the optimal execution locations for each \oran application, thus moving network intelligence to the edge with up to 2.6$\times$ reduction of control overhead over the \oran E2 interface. To the best of our knowledge, this is the first large-scale demonstration of an \oran-compliant network intelligence orchestration system.


\vspace{-0.1cm}
\section{Related Work} \label{sec:related}

The application of ML/AI algorithms to cellular networks is gaining momentum as a promising and effective way to design and deploy solutions capable of predicting,  controlling, and automating the network behavior under dynamic conditions. Relevant examples include the application of Deep Learning and \gls{drl} to predict the network load~\cite{polese2020machine, bega2019deepcog,salhab2021autonomous}, classify traffic~\cite{li2021robust,paul2019traffic,weerasinghe2019supervised}, perform beam alignment~\cite{polese2021deepbeam,klautau20185g}, allocate radio resources~\cite{9206115, bonati2021scope, 9468707}, and deploy service-tailored network slices~\cite{okic2020pi,bonati2021intelligence,bega2020network,9490293,salhab2021autonomous,9459763,chergui2020opex}.
It is clear that ML/AI techniques will play a key role in the transition to intelligent networks, especially in the \oran ecosystem~\cite{lee2020hosting}.
However, a relevant challenge that still remains unsolved is how to bring such intelligence to the network in an efficient, reliable and automated way, which is ultimately the goal of this paper.

In \cite{ayalainfocom2021}, Ayala-Romero et al.\ present an online Bayesian learning orchestration framework for intelligent virtualized \glspl{ran} where resource allocation follow channel conditions and network load.
The same authors present a similar framework in \cite{9286741}, where networking and computational resources are orchestrated via \gls{drl} to comply with \glspl{sla} while accounting for the limited amount of resources.
Singh et al.\ present GreenRAN, an energy-efficient orchestration framework for NextG that splits and allocates \gls{ran} components according to the current resource availability~\cite{singh2020energy}.
%
%
In \cite{9489294}, Chatterjee et al. present a radio resource orchestration framework for 5G applications where network slices are dynamically re-assigned to avoid inefficiencies and \gls{sla} violations.
Relevant to our work are the works of Morais et al.~\cite{morais2021placeran} and Matoussi et al.~\cite{matoussi20205g}, which present frameworks to optimally disaggregate, place and orchestrate \gls{ran} components in the network to minimize computation and energy consumption while accounting for diverse latency and performance requirements.
Although the above works all present orchestration frameworks for NextG systems, they are focused on orchestrating \gls{ran} resources and functionalities, rather than network intelligence, which represents a substantially different problem.

In the context of orchestrating ML/AI models in NextG systems, Baranda et al.~\cite{baranda2020integration,9484610}
present an architecture for the automated deployment of models in the 5Growth \gls{mano} platform~\cite{li20215growth}, and demonstrate
 automated instantiation of models on demand.
The closest to our work is the work of Salem et al.~\cite{salem2021towards}, which proposes an orchestrator to select and instantiate inference models at different locations of the network to obtain a desirable balance between accuracy and latency. However, \cite{salem2021towards} is not concerned with \oran systems, but focuses on data-driven solutions for inference in cloud-based applications.

Besides the differences highlighted in the previous discussion, \name differs from the above works in that it focuses on the Open \gls{ran} architecture and is designed to instantiate \textit{both} inference and control solutions complying with \oran specifications.
Moreover, \name allows model sharing across multiple requests to efficiently reuse available network resources. We prototyped and benchmarked \name on Colosseum. To the best of our knowledge, this is the first large-scale demonstration of a network intelligence orchestration system tailored to \oran architecture and networks.

\section{O-RAN: A Primer} \label{sec:primer}

%
\oran embraces the 7-2x functional split (an extension of the \gls{3gpp} 7-2 split), where network functionalities are divided across multiple nodes, namely, \glspl{cu}, \glspl{du} and \glspl{ru} (Fig.~\ref{fig:system_split}, left). The \glspl{ru} implement lower physical layer functionalities. The \glspl{du} interact with the \glspl{ru} via the Open Fronthaul interface and implement functionalities pertaining to both the higher physical layer and the \gls{mac} layer. Finally, the remaining functionalities of the protocol stack are implemented and executed in the \gls{cu}. The latter is connected to the \glspl{du} through the F1 interface and is further split in two entities---handling control and user planes---connected via the E1 interface.
%
These network elements run on ``white-box'' hardware components connected through O-RAN open interfaces, thus enabling multivendor interoperability and overcoming the vendor lock-in~\cite{bonati2020open}.

Beyond disaggregation, the main innovation introduced by \oran lies in the \nonrt and \nearrt \glspl{ric}.
These components enable dynamic and softwarized control of the \gls{ran}, as well as the collection of statistics via a publish-subscribe model~\cite{oran-e2sm} through open and standardized interfaces, e.g., the O1 and E2 interfaces (Fig.~\ref{fig:system_split}, left).
Specifically, the \nearrt \gls{ric} hosts applications (\textit{xApps}) that implement time-sensitive---i.e., between 10~ms and 1~s---operations to perform closed-loop control over the \gls{ran} elements. Practical examples include control of load balancing, handover procedures, scheduling and \gls{ran} slicing policies~\cite{bonati2021intelligence,doro2021coordinated,doro2020slicing,doro2020sledge}.
The \nonrt \gls{ric}, instead, is designed to execute within a \gls{smo} framework, e.g., \gls{onap}, and acts at time scales above 1~s. It takes care of training \gls{ml}/\gls{ai} models, as well as deploying models and network control policies on the \nearrt \gls{ric} through the A1 interface. Similar to its \nearrt counterpart, the \nonrt \gls{ric} supports the execution of third-party applications, called \textit{rApps}.
These components act in concert to gather data and performance metrics from the \gls{ran}, and to optimize and reprogram its behavior in real time through software algorithms to reach \gls{no}'s goals.
\oran specifications also envision \gls{ml}/\gls{ai} models instantiated directly on the \glspl{cu} and \glspl{du}, implementing \rt---\gls{tti} level---control loops that operate on 10~ms time-scales~\cite{oran-ml}.
%
Although these are left for future \oran extensions, \name has been natively designed to  support such control loops, implementing \rt applications, which we will refer to as \textit{dApps} to avoid confusion.

\section{\name}
\label{sec:system}

\begin{figure}[t]
\begin{centering}
    \includegraphics[width=0.82\columnwidth]{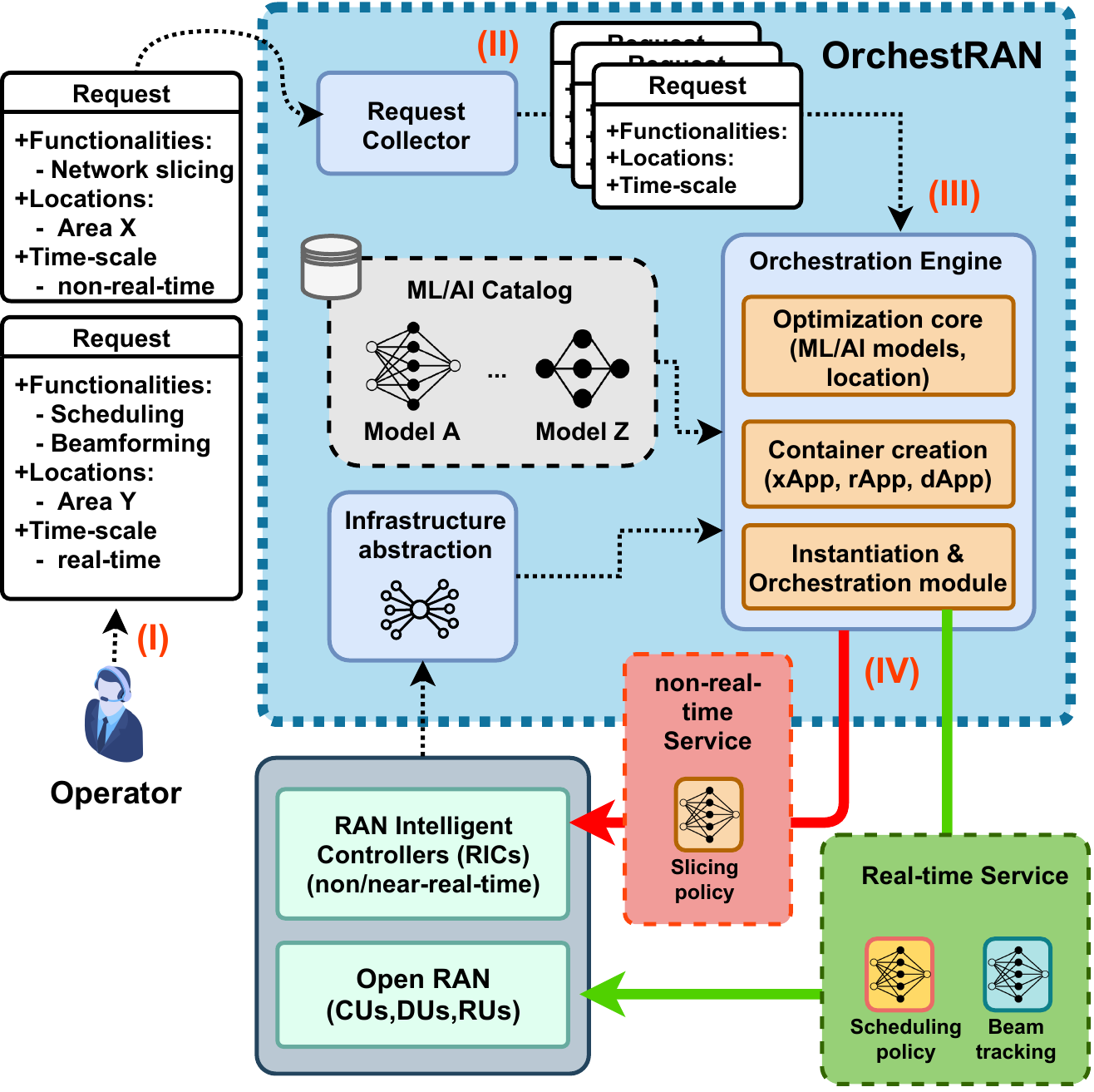}
    \vspace{-0.4cm}
    \caption{System design of \name and main procedures.}
    \label{fig:system}
\end{centering}
\end{figure}

As illustrated in Fig.~\ref{fig:system_split}, \name is  designed to be executed as an \textit{rApp} at the \nonrt \gls{ric}. Its architecture is illustrated in Fig. \ref{fig:system}.
%
%
At a high-level, first \glspl{no} specify their intent by submitting a control request to \name (step I). This includes the set of functionalities they want to deploy (e.g., network slicing, beamforming, scheduling control, etc.), the location where functionalities are to be executed (e.g., \gls{ric}, \gls{cu}, \gls{du}) and the desired time constraint (e.g., delay-tolerant, low-latency).
Then, requests are gathered by the \textit{Request Collector} (step II, Section~\ref{sec:control_request}) and fed to the \textit{Orchestration Engine} (step III, Section~\ref{sec:orchestration_engine}) which: (i)~Accesses the \textit{ML/AI Catalog} (Section~\ref{sec:catalog}) and the \textit{Infrastructure Abstraction} module (Section~\ref{sec:infra}) to determine the optimal orchestration policy and models to be instantiated; (ii)~automatically creates containers with the embedded ML/AI models in the form of \oran applications,
and (iii)~ dispatches such applications at the locations determined by the Orchestration Engine.

\subsection{The Infrastructure Abstraction Module}
\label{sec:infra}
\vspace{-0.1cm}
This module provides a high-level representation of the physical \gls{ran} architecture, which is
divided into five separate logical groups: \nonrt \glspl{ric}, \nearrt \glspl{ric}, \glspl{cu}, \glspl{du} and \glspl{ru}. Each group contains a different number of nodes
deployed at different locations of the network.
Let $\devs$ be the set of such nodes, and $D=|\devs|$ be their number.

The hierarchical relationships between nodes
can be represented via an undirected graph with a tree structure such as the one in Fig.~\ref{fig:system_split} (right). Specifically, leaves represent nodes at the edge (e.g., \glspl{ru}/\glspl{du}/\glspl{cu}), while the \nonrt \gls{ric} is the root of the tree.\footnote{Coexisting \glspl{cu}/\glspl{du}/\glspl{ru} are modeled as a single logical node with a hierarchy level equal to that of the hierarchically highest node in the group.}
%
For any two nodes $d', d'' \in \devs$, we define variable $c_{d',d''}\in\{0,1\}$ such that $c_{d',d''}=1$ if node $d'$ is reachable from node $d''$ (e.g., there exist a communication link such that node $d'$ can forward data to node $d''$), $c_{d',d''}=0$ otherwise. In practical deployments, it is reasonable to assume that nodes on different branches of the tree are unreachable.
Moreover, for each node $d\in\devs$, let $\rho^\xi_d$ be the total amount of resources of type $\xi\in\Xi$ dedicated to hosting and executing ML/AI models and their functionalities, where $\Xi$ represents the set of all resource types. Although we do not make any assumptions on the specific types of resources, practical examples may include the number of CPUs, GPUs, as well as available disk storage and memory. In the following, we assume that each \nonrt \gls{ric} identifies an independent networking domain and the set of nodes $\devs$ includes \nearrt \glspl{ric}, \glspl{cu}, \glspl{du} and \glspl{ru} controlled by the corresponding \nonrt \gls{ric} only.


\subsection{The ML/AI Catalog} \label{sec:catalog}
\vspace{-0.1cm}
In \name, the available pre-trained data-driven solutions are stored in a \textit{ML/AI Catalog} consisting of a set $\mods$ of ML/AI models.
Let $\funcs$ be the set of all possible control and inference functionalities (e.g., scheduling, beamforming, capacity forecasting, handover prediction) offered by such ML/AI models---hereafter referred to simply as ``models".

Let $M=|\mods|$ and $F=|\funcs|$. For each model $m\in\mods$, $\funcs_m \subseteq \funcs$ represents the subset of functionalities offered by $m$. Accordingly, we define a binary variable $\sigma_{m,f}\in\{0,1\}$ such that $\sigma_{m,f}=1$ if $f\in\funcs_m$, $\sigma_{m,f}=0$ otherwise.
%
We use $\rho^\xi_m$ to indicate the amount of resources of type $\xi\in\Xi$ required to instantiate and execute model $m$. Let $\types$ be the set of possible input types. For each model $m\in\mods$, $\tin_m\in\types$ represents the type of input required by the model (e.g., IQ samples, throughput and buffer size measurements).

Naturally, not all models can be equally executed everywhere. For example, a model $m$ performing beam alignment~\cite{polese2021deepbeam}, in which received IQ samples are fed to a neural network to determine the beam direction, can only execute on nodes where IQ samples are available. While IQ samples can be accessed in real-time at the \gls{ru}, they are unlikely to be available at \glspl{cu} and the \glspl{ric} without incurring in
high overhead and transmission latency.
For this reason, we introduce a suitability indicator $\beta_{m,f,d}\in[0,1]$ which specifies how well a model $m$ is suited to provide a specific functionality $f\in\funcs$ when instantiated on node $d$. Values of $\beta_{m,f,d}$ closer to 1 mean that the model is well-suited to execute at a specific location, while values closer to 0 indicate that the model performs poorly.
We also introduce a performance score $\gamma_{m,f}$ measuring the performance of the model with respect to $f\in\funcs$. Typical performance metrics include classification/forecasting accuracy, mean squared error and probability of false alarm.
A model can be instantiated on the same node multiple times to serve different \glspl{no} or traffic classes. However,
due to limited resources, each node $d$ supports at most $C_{m,d}=\min_{\xi\in\Xi} \{\lfloor \rho^\xi_d / \rho^\xi_m \rfloor\}$ instances of model $m$,
where $\lfloor \cdot \rfloor$ is the floor operator.

\subsection{Request Collector} \label{sec:control_request}
\vspace{-0.1cm}
\name allows \glspl{no} to submit requests specifying which functionalities they require, where they should execute, and the desired performance and timing requirements. Without loss of generality, we assume that each request is feasible.
The Request Collector of \name is in charge of collecting such requests. A request $i$ is defined as a tuple $(\funcs_i,\boldsymbol{\pi}_i,\boldsymbol{\delta}_{i}, \devs^{\mathrm{IN}}_{i})$, with each element defined as follows:

\noindent
$\bullet$ \textit{Functions and locations.} For each request $i$, we define the set of functionalities that must be instantiated on the nodes as $\funcs_i=(\funcs_{i,d})_{d\in\devs}$, with $\funcs_{i,d}\subseteq\funcs$. Required functionalities and nodes are specified by a binary indicator $\tau_{i,f,d}\in\{0,1\}$ such that $\tau_{i,f,d}=1$ if request $i$ requires functionality $f$ on node $d$, i.e.,  $f\in\funcs_{i,d}$, $\tau_{i,f,d}=0$ otherwise. We also define $\devs_i=\{d\in\devs: \sum_{f\in\funcs_i} \tau_{i,f,d} \geq 1\}$ as the subset of nodes of the network where functionalities in $\funcs_i$ should be offered;

\noindent
$\bullet$ \textit{Performance requirements.} For any request $i$, $\boldsymbol{\pi}_i=(\pi_{i,f,d})_{d\in\devs_i, f\in\funcs_{i,d}}$ indicates the minimum performance requirements that must be satisfied to accommodate $i$. For example, if $f$ is a beam detection functionality, $\pi_{i,f,d}$ can represent the minimum detection accuracy of the model.
We do not make any assumptions on the physical meaning of $\pi_{i,f,d}$ as it reasonably differs from one functionality to the other.

\noindent
$\bullet$ \textit{Timing requirements.}
Some functionalities might have strict latency requirements that make their execution at nodes far away from the location where the input is generated impractical or inefficient. For this reason, $\delta_{i,f,d}\geq0$ represents the maximum latency request $i$ can tolerate in executing $f$ on $d$;

\noindent
$\bullet$ \textit{Data source.} For each request $i$, the \gls{no} also specifies the subset of nodes whose generated (or collected) data must be used to deliver functionality $f$ on node $d$. This set is defined as $\devs^{\mathrm{IN}}_{i}=(\devs^{\mathrm{IN}}_{i,f,d})_{d\in\devs_i, f\in\funcs_{i,d}}$, where $\devs^{\mathrm{IN}}_{i,f,d}\subseteq\devs$. This information is paramount to ensure that each model is fed with the proper data generated by the intended sources only. For any tuple $(i,f,d)$ we assume that $c_{d,d'}=1$ for all $d'\in\devs^{\mathrm{IN}}_{i,f,d}$.

In the remaining of this paper, we use $\reqs$ to represent the set of outstanding requests with $I=|\reqs|$ being their number.

\vspace{-0.1cm}
\subsection{The Orchestration Engine} \label{sec:orchestration_engine}
\begin{figure}[t]
\setlength\abovecaptionskip{-.25cm}
\begin{centering}
    \includegraphics[width=0.9\columnwidth]{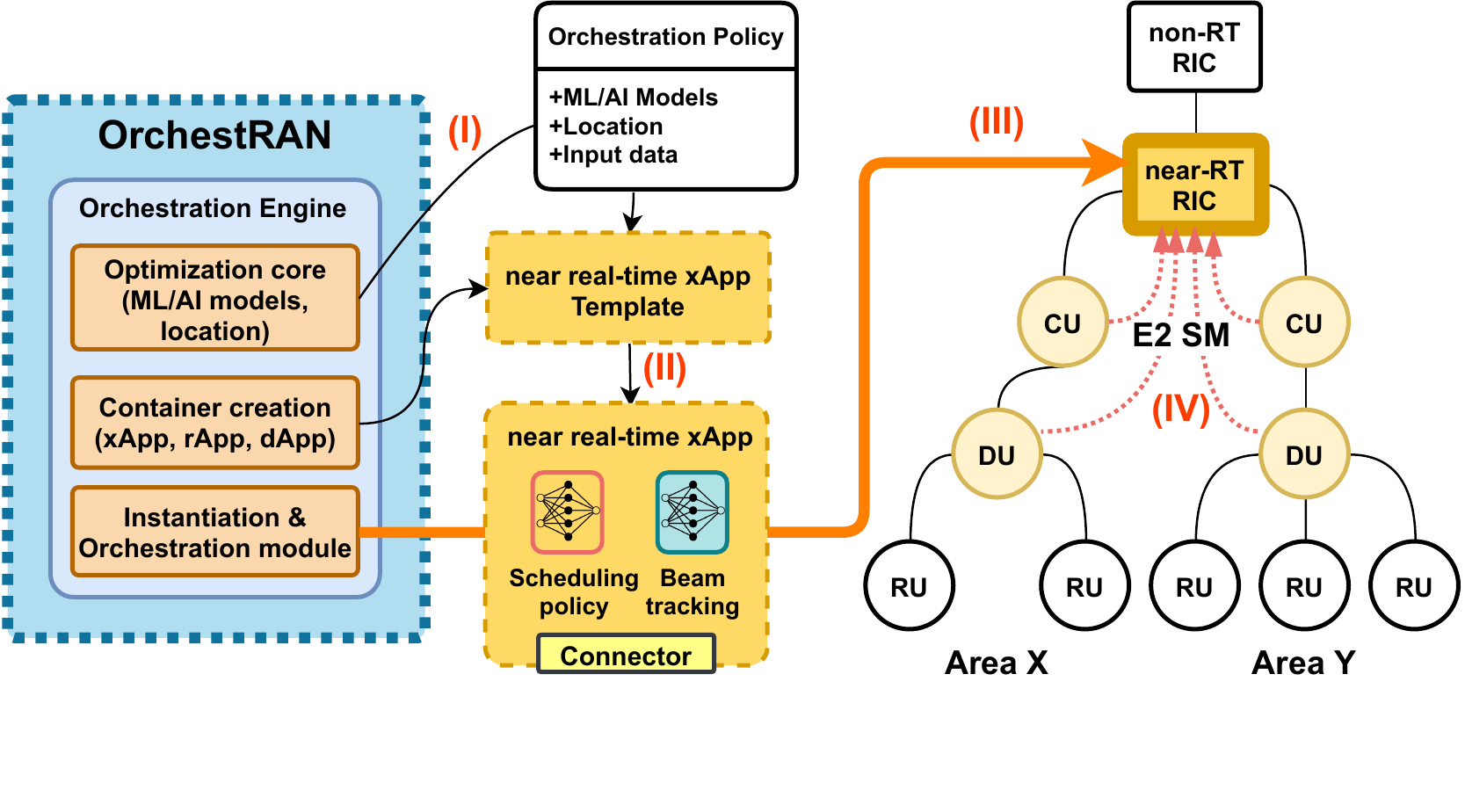}
    \vspace{-0.4cm}
    \caption{An example of creation and dispatchment of an xApp on the \nearrt \gls{ric} via \name.}
    \label{fig:dispatch}
\end{centering}
\end{figure}

\vspace{-0.1cm}
As depicted in Fig.~\ref{fig:dispatch}, once requests are submitted to \name,
the Orchestration Engine selects the most suitable models from the ML/AI Catalog and the location where they should execute (step I). Then, \name embeds the models into containers (e.g., Docker containers of dApps, xApps, rApps) (step II) and dispatches them to the selected nodes (step III). Here, they are fed data from the \gls{ran} and execute their functionalities (step IV).
The selection of the models and of their optimal execution location is performed by solving the \textit{orchestration problem} discussed in detail in Sections~\ref{sec:orchestration_problem} and~\ref{sec:solutions}.
This results in an \textit{orchestration policy}, which is converted into a set of \oran applications that are dispatched and executed at the designated network nodes, as discussed next.

\noindent
\textit{Container creation, dispatchment and instantiation.}
%
%
To embed models in different \oran applications, containers integrate two subsystems, which are automatically compiled from descriptive files upon instantiation. The first is the model itself, and the second is an application-specific \textit{connector}. This is a library that interfaces with the node where the application is running (i.e., with the \gls{du}
in the case of dApps, \nearrt \gls{ric} for xApps, and \nonrt \gls{ric} for rApps), collects data from $\devs^{\mathrm{IN}}_{i}$ and sends control commands to nodes in $\devs_i$.
%
Once the containers are generated, \name dispatches them to the proper endpoints specified in the orchestration policy, where are instantiated and interfaced with the \gls{ran} to receive input data. For example, xApps automatically send an E2 subscription request to nodes in $\devs^{\mathrm{IN}}_{i}$, and use custom \glspl{sm} to interact with them over the E2 interface~\cite{oran-e2sm} (see Fig.~\ref{fig:dispatch}).

\section{The Orchestration Problem} \label{sec:orchestration_problem}



Before formulating the orchestration problem, we first discuss important properties of Open RAN systems.

\noindent
$\bullet$~\textit{Functionality outsourcing.}
Any functionality that was originally intended to execute at node $d'$ can be outsourced to any other node $d''\in\devs$ as long as $c_{d',d''} = 1$.
As we will discuss next, the node hosting the outsourced model must have access to the required input data, have enough resources to instantiate and execute the outsourced model, and must satisfy performance and timing requirements of the original request.

\noindent
$\bullet$~\textit{Model sharing.} The limited amount of resources, especially at \glspl{du} and \glspl{ru}, calls for efficient resource allocation strategies. If multiple requests involve the same functionalities on the same group of nodes, an efficient approach consists in deploying a single model that can be shared across all requests.

\begin{figure}[t]
\begin{centering}
    \includegraphics[width=0.75\columnwidth]{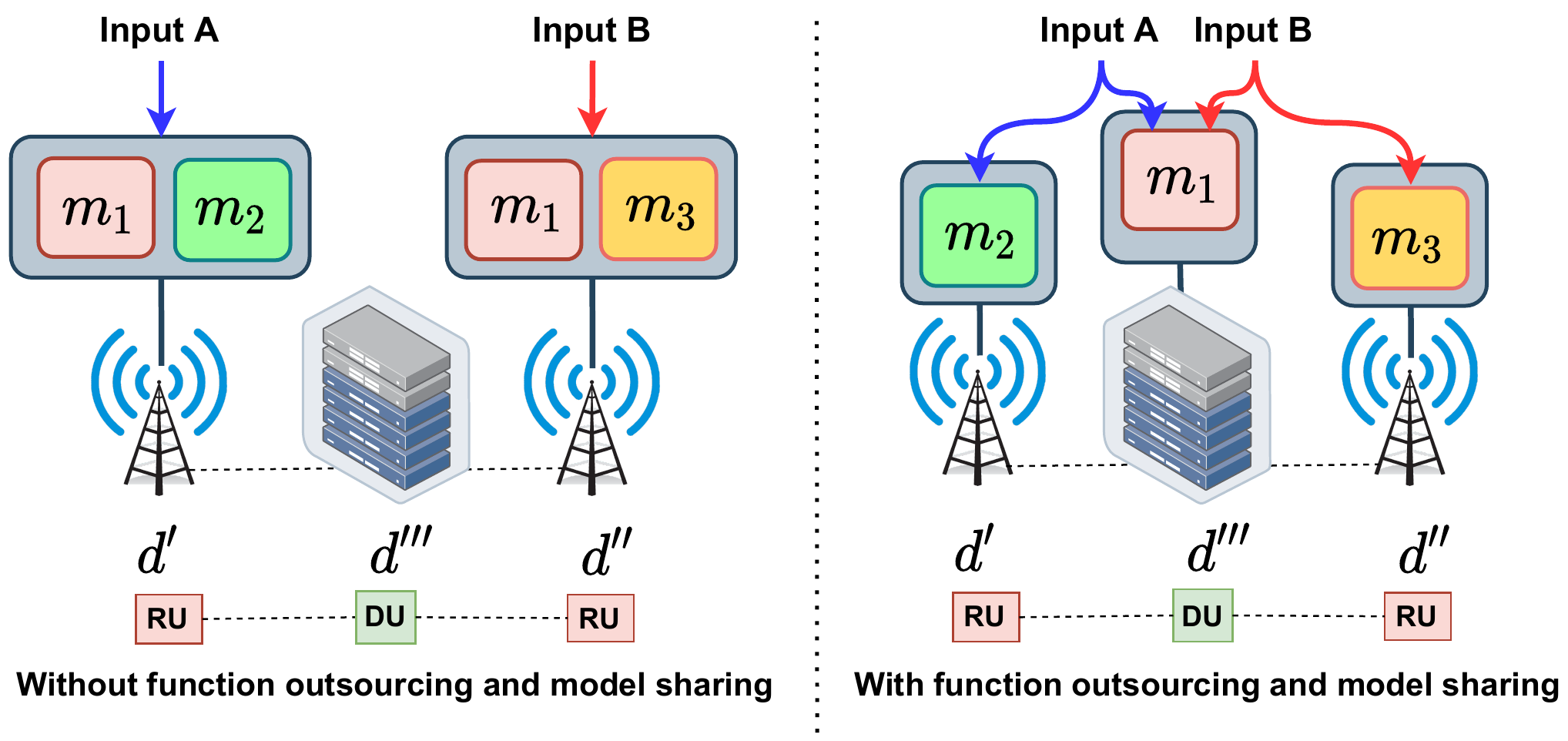}
    \vspace{-0.4cm}
    \caption{Example of function outsourcing and model sharing in Open RAN.}
    \label{fig:sharing}
\end{centering}
\end{figure}

For the sake of clarity,
in Fig.~\ref{fig:sharing} (left) we show an example
where a request can be satisfied by instantiating models $m_1$ and $m_2$ on $d'$, and a second one that can be accommodated by instantiating models $m_1$ and $m_3$ on $d''$.
Fig.~\ref{fig:sharing} (right) shows an alternative solution where $m_1$ (common to both requests) is \textit{outsourced} to $d'''$ and it is shared between the two requests, with a total of three deployed models, against the four required in Fig.~\ref{fig:sharing} (left).
%
In the next section, we also discuss the case where model sharing or function outsourcing are nonviable.


\subsection{Formulating the Orchestration Problem} \label{sec:formulation}
\vspace{-0.1cm}
Let $x^{i,f,d}_{m,k,d'}\in\{0,1\}$ be a binary variable such that $x^{i,f,d}_{m,k,d'}=1$ if functionality $f$ demanded by request $i$ on node $d$ is provided by instance $k$ of model $m$ instantiated on node $d'$. In the following, we refer to the variable $\mathbf{x} = (x^{i,f,d}_{m,k,d'})_{i,f,d,m,k,d'}$ as the \textit{orchestration policy}, where $i\!\in\!\reqs, f\!\in\!\funcs, (d,d')\!\in\!\devs\!\times\!\devs,m\!\in\!\mods,k\!=\!1\dots C_{m,d'}$.

\noindent
$\bullet$~\textit{Conflict avoidance.}~For any tuple $(i,f,d)$ such that $\tau_{i,f,d}\!=\!1$, we assume that \name can instantiate at most one model to avoid multiple models controlling the same parameters and/or functionalities.
As mentioned earlier, this can be achieved by either instantiating the model at $d$, or by outsourcing it to another node $d'\neq d$.
The above requirement can be formalized as follows:
\vspace{-0.1cm}
\begin{equation}
    \sum_{m\in\mods}\sigma_{m,f} \sum_{d'\in\devs} \sum_{k=1}^{C_{m,d'}} c_{d,d'} x^{i,f,d}_{m,k,d'} = y_i \tau_{i,f,d} \label{eq:con1}
\end{equation}
\noindent
where $y_i\in\{0,1\}$ indicates whether or not $i$ is satisfied. Specifically, \eqref{eq:con1} ensures that: (i)~For any tuple $(i,f,d)$ such that $\tau_{i,f,d}=1$, function $f$ is provided by one model only, and (ii) $y_i=1$ (i.e., request $i$ is satisfied) if and only if \name deploys models providing all functionalities specified in $\funcs_i$.

\noindent
$\bullet$~\textit{Complying with the requirements.}~An important aspect of the orchestration problem is guaranteeing that the orchestration policy $\mathbf{x}$ satisfies the minimum performance requirements $\boldsymbol{\pi}_i$ of each request $i$, and that both data collection and execution procedures do not exceed the maximum latency constraint $\delta_{i,f,d}$.
%
These requirements are captured by the following constraints.

\subsubsection{Quality of models}
For each tuple $(i,f,d)$ such that $\tau_{i,f,d}=1$, \glspl{no} can specify a minimum performance level $\pi_{i,f,d}$. This can be enforced via the following constraint
\vspace{-0.1cm}
\begin{align}
   \chi_{i,f,d} \!\!\sum_{m\in\mods} \sum_{d'\in\devs} \sum_{k=1}^{C_{m,d'}} c_{d,d'} x^{i,f,d}_{m,k,d'} A_{m,f,d}
    \geq \chi_{i,f,d} y_i \pi_{i,f,d}  \label{eq:con3}
\end{align}
\noindent
where $A_{m,f,d} = \beta_{m,f,d} \, \gamma_{m,f} \, \sigma_{m,f}$, and the performance score $\gamma_{m,f}$ is defined in Section \ref{sec:catalog}. In \eqref{eq:con3}, $\chi_{i,f,d}=1$ if the goal is to guarantee a value of $\gamma_{m,f}$ higher than a minimum performance level $\pi_{i,f,d}$, and $\chi_{i,f,d} = -1$ if the goal is to keep $\gamma_{m,f}$ below a maximum value $\pi_{i,f,d}$.


\subsubsection{Control-loop time-scales}


Each model $m$ requires a specific type of input $\tin_m$ and,
for each tuple $(i,f,d)$, we must ensure that the time needed to collect such input from nodes in $\devs^{\mathrm{IN}}_{i,f,d}$ does not exceed $\delta_{i,f,d}$.
For each orchestration policy $\mathbf{x}$, the \textit{data collection time} can be formalized as follows:
\vspace{-0.1cm}
\begin{align}
\Delta_{i,f,d}(\mathbf{x}) =
\!\!\sum_{m\in\mods}\!\! \sigma_{m,f}\!\! \sum_{d'\in\devs}\!\! \sum_{k=1}^{C_{md'}}\!\! x^{i,f,d}_{m,k,d'}\!\!\!\!\!\!\! \sum_{d''\in\devs^{\mathrm{IN}}_{i,f,d}} \!\!\!\!\!c_{d',d''} \Theta^{i,f,d}_{m,d',d''} \label{eq:delta_d}
\end{align}
\noindent
where $\Theta^{i,f,d}_{m,d',d''} = \left(\frac{s_{\tin_m}}{b_{d',d''} |\devs^{\mathrm{IN}}_{i,f,d}|}\! + T_{d',d''} \right)$, $s_{\tin_m}$ is the input size of model $m$ measured in bytes, $b_{d',d''}$ is the data rate of the link between nodes $d''$ and $d'$, and
$T_{d',d''}$ represents the propagation delay between nodes $d'$ and $d''$.
Let $T^{exec}_m$ be the time to execute model $m$ on node $d'$. For any tuple $(i,f,d)$, the \textit{execution time} under orchestration policy $\mathbf{x}$ is
\vspace{-0.2cm}
\begin{equation}
    \Delta^{EXEC}_{i,f,d}(\mathbf{x}) = \sum_{m\in\mods}\sigma_{m,f} \sum_{d'\in\devs} T^{exec}_{m,d'} \sum_{k=1}^{C_{m,d'}} x^{i,f,d}_{m,k,d'} \label{eq:execution}
\end{equation}

By combining \eqref{eq:delta_d} and \eqref{eq:execution}, any orchestration policy $\mathbf{x}$ must satisfy the following constraint for all $(i,f,d)$ tuples:
\begin{equation}
  \Delta_{i,f,d}(\mathbf{x}) + \Delta^{EXEC}_{i,f,d}(\mathbf{x}) \leq \delta_{i,f,d} \; \tau_{i,f,d} \label{eq:latency_tot}
\end{equation}

\noindent
$\bullet$~\textit{Avoiding resource over-provisioning.}~We must guarantee that the resources consumed by the \oran applications do not exceed the resources $\rho^\xi_{d}$ of type $\xi$ available at each node (i.e., $\rho^\xi_{d}$). For each $d\in\devs$ and $\xi\in\Xi$, we have
\vspace{-0.1cm}
\begin{equation}
    \sum_{m\in\mods} \rho^\xi_{m} \sum_{k=1}^{C_{md}} z_{m,k,d} \leq \rho^\xi_{d} \label{eq:con5}
\end{equation}
\noindent
where $z_{m,k,d}\in\{0,1\}$ indicates whether instance $k$ of model $m$ is associated to at least one model on node $d$. Specifically, let
\vspace{-0.1cm}
\begin{equation}
    n_{m,k,d} = \sum_{i\in\reqs} \sum_{f\in\funcs_i} \sum_{d'\in\devs} x^{i,f,d'}_{m,k,d} \label{eq:n}
\end{equation}
\noindent
be the number of tuples $(i,f,d')$ assigned to instance $k$ of model $m$ on node $d$
($n_{m,k,d}>1$ implies that $m$ is shared). Notice that \eqref{eq:con5} and \eqref{eq:n} are coupled one to another as $z_{m,k,d}=1$ if and only if $n_{m,k,d}>0$.
This conditional relationship can be formulated by using the following big-M formulation~\cite{RAMAN1994563}
\vspace{-0.1cm}
\begin{align}
    n_{m,k,d} & \geq 1 - M ( 1- z_{m,k,d}) \label{eq:aux:1} \\
    n_{m,k,d} & \leq M z_{m,k,d} \label{eq:aux:2}
\end{align}
\noindent
where $M\in\mathbb{R}$ is a real-valued number whose value is larger than the maximum value of $n_{m,k,d}$, i.e., $M>IFD$~\cite{RAMAN1994563}.

\noindent
$\bullet$~\textit{Problem formulation.}~For any request $i$, let $v_i\geq0$ represent its value. The goal of \name is to compute an orchestration policy $\mathbf{x}$ maximizing the total value of requests being accommodated by selecting (i) which requests can be accommodated; (ii) which models should be instantiated; and (iii) where they should be executed to satisfy request performance and timescale requirements. This can be formulated as
\vspace{-0.1cm}
   \begin{align}
   \max_{\mathbf{x},\mathbf{y},\mathbf{z}} \hspace{1cm} & \sum_{i\in\reqs} y_i v_i &\label{eq:problem} \\
   \mathrm{subject~to} \hspace{1cm} & \mbox{Constraints}~ \eqref{eq:con1},
   \eqref{eq:con3},  \eqref{eq:latency_tot}, \eqref{eq:con5}, \eqref{eq:aux:1}, \eqref{eq:aux:2} \nonumber \\
   \hspace{1cm} & x^{i,f,d}_{m,k,d'} \in \{0,1\} \\
   \hspace{1cm} & y_{i}\in\{0,1\} \\
   \hspace{1cm} & z_{m,k,d}\in\{0,1\}
   \end{align}
\noindent
where $\mathbf{x}$ is the orchestration policy, $\mathbf{y}=(y_{i})_{i\in\reqs}$ and $\mathbf{z}=(z_{m,k,d})_{m\in\mods,k=1,\dots,C_{m,d},d\in\devs}$.
A particularly relevant case is that where $v_i=1$ for all $i\in\reqs$, i.e., the goal of \name is to maximize the number of satisfied requests.
%

\noindent
$\bullet$~\textit{Disabling model sharing.}
Indeed, model sharing allows a more efficient use of the available resources.
However, out of privacy and business concerns, \glspl{no} might not be willing to share \oran applications. In this case, model sharing can be disabled in \name by guaranteeing that a model is assigned to one request only. This is achieved by adding the following constraint for any $m\in\mods$, $d'\in\devs$ and $k=1,..,C_{m,d'}$
\vspace{-0.1cm}
\begin{equation} \label{eq:no_sharing}
    \sum_{i\in\reqs} \sum_{d\in\devs} \sum_{f\in\funcs_{i,d}} x^{i,f,d}_{m,k,d'} \leq 1
\end{equation}

\subsection{NP-hardness of the Orchestration Problem} \label{sec:nphard}
\vspace{-0.1cm}
Problem~\eqref{eq:problem} is a Binary Integer Linear Programming (BILP) problem which can be shown to be NP-hard. The proof consists in building a polynomial-time reduction of the 3-$\mathrm{SAT}$ problem (which is NP-complete) to an instance of Problem \eqref{eq:problem}~\cite{karp1972reducibility}.

\section{Solving the Orchestration Problem} \label{sec:solutions}

BILP problems such as Problem \eqref{eq:problem} can be optimally solved via Branch-and-Bound (B\&B) techniques~\cite{wolsey2020integer}, 
readily available within well-established numerical solvers, e.g., CPLEX, MATLAB, Gurobi.
However, due to the extremely large number $N_{\mathrm{OPT}}$ of optimization variables, these solvers might still fail to compute an optimal solution in a reasonable amount of time, especially in large-scale deployments.
Indeed, $N_{\mathrm{OPT}}\!=\!|\mathbf{x}|\! +\! |\mathbf{y}| \!+\! |\mathbf{z}|\! \approx\! |\mathbf{x}|$, where $|\mathbf{x}| \!=\! \mathcal{O}(IFD^2MC_{\mathrm{max}})$, $|\mathbf{y}| \!=\! \mathcal{O}(I)$, $|\mathbf{z}| \!=\! \mathcal{O}(MDC_{\mathrm{max}})$, and
$C_{\mathrm{max}}\!=\!\max_{m\in\mods,d\in\devs}\{C_{m,d}\}$.
For example, a deployment with $D=20$, $M=13$, $I=10$, $F=7$ and $C_{\mathrm{max}}=3$ involves $\approx$10$^6$ optimization variables.


\subsection{Combating Dimensionality via Variable Reduction} \label{sec:variable_reduction}
\vspace{-0.1cm}
To mitigate the ``curse of dimensionality'' of the orchestration problem, we have developed two pre-processing algorithms to reduce the complexity of Problem~\eqref{eq:problem} while guaranteeing the optimality of the computed solutions. We leverage a technique called \textit{variable reduction}~\cite{8770156}. This exploits the fact that, due to constraints and structural properties of the problem, there might exist a subset of \textit{inactive} variables whose value is always zero.
These variables do not participate in the optimization process, yet they increase its complexity. To identify those variables, we have designed the following two techniques.

\noindent
$\bullet$ \textit{Function-aware Pruning (FP).} It identifies the set of inactive variables $\mathbf{x}^{\mathrm{FP}}_{-} =  \{x^{i,f,d}_{m,k,d'} : \tau_{i,f,d} = 0 \lor \sigma_{m,f} = 0, \forall i\in \reqs, f\in\funcs, (d,d')\in\devs\times\devs, m\in\mods, k=1,\dots,C_{,m,d}\}$,
which contains all the $x^{i,f,d}_{m,k,d}$ variables such that either (i) $\tau_{i,f,d}=0$, i.e., request $i$ does not require function $f$ at node $d$, or (ii) $\sigma_{m,f}=0$, i.e., model $m$ does not offer function $f$;

\noindent
$\bullet$ \textit{Architecture-aware Pruning (AP).} This procedure identities those variables whose activation results in instantiating a model on a node that cannot receive input data from nodes in $\devs^{\mathrm{IN}}_{i,f,d}$. Indeed, for a given tuple $(i,f,d)$ such that $\tau_{i,f,d}=1$, we cannot instantiate any model on a node $d'$ such that $c_{d,d'}=0$, i.e., the two nodes are not connected.
%
The set of these inactive variables is defined as
$\mathbf{x}^{\mathrm{AP}}_{-} =  \{x^{i,f,d}_{m,k,d'} : c_{d,d'} = 0, \forall i\in \reqs, f\in\funcs,(d,d')\in\devs\times\devs, m\in\mods, k=1,\dots,C_{,m,d}\}$.

Once we have identified all inactive variables,
Problem \eqref{eq:problem} is cast into a lower-dimensional space where the new set of optimization variables is equal to $\tilde{\mathbf{x}} = \mathbf{x} \setminus \{\mathbf{x}^{\mathrm{FP}}_{-} \cup \mathbf{x}^{\mathrm{AP}}_{-}\}$, which still guarantees the optimality of the solution~\cite{8770156}.
%
%
The impact of these procedures on the complexity of the orchestration problem will be investigated in Section \ref{sec:numerical}.

\subsection{Graph Tree Branching} \label{sec:branching}
\vspace{-0.1cm}
Notice that $|\mathbf{x}| = \mathcal{O}(IFD^2MC_{\mathrm{max}})$, i.e., the number of variables of the orchestration problem grows quadratically in the number $D$ of nodes.
%
Since the majority of nodes of the infrastructure are \glspl{ru}, \glspl{du} and \glspl{cu}, it is reasonable to conclude that these nodes are the major source of complexity.
Moreover,
\oran systems operate following a cluster-based approach where each \nearrt \gls{ric} controls a subset of \glspl{cu}, \glspl{du} and \glspl{ru} of the network only, i.e., a \textit{cluster}, which have none (or limited) interactions with nodes from other clusters.

These two intuitions are the rationale behind the low-complexity and scalable solution proposed in this section, which consists in splitting the infrastructure tree into smaller subtrees---each operating as an individual cluster---and creating sub-instances of the orchestration problem that only accounts for requests and nodes regarding the considered subtree.
%
The main steps of this algorithm are:

\noindent
$\bullet$ \textit{Step I}: Let $C$ be the number of \nearrt \glspl{ric} in the \nonrt \gls{ric} domain. For each cluster $c$, the $c$-th subtree $\devs_c\subseteq\devs$ is defined such that $\devs = \bigcup_{c=1}^C\devs_c$ and $\bigcap_{c=1}^C \devs_c = d^{\mathrm{root}}$, with $d^{\mathrm{root}}$ being the \nonrt \gls{ric}. A variable $\alpha_{d,c}\in\{0,1\}$ is used to determine whether a node $d\in\devs$ belongs to cluster $c$ (i.e., $\alpha_{d,c}=1$) or not (i.e., $\alpha_{d,c}=0$).
Since, $\bigcap_{c=1}^C \devs_c = d^{\mathrm{root}}$, we have that $\sum_{c=1}^C \alpha_{d,c}=1$ for any $d\in\devs\setminus\{d^{\mathrm{root}}\}$;

\noindent
$\bullet$ \textit{Step II}: For each subtree $\devs_c$ we identify the subset $\reqs_c\subseteq\reqs$ such that $\reqs_c=\{i\in\reqs: \sum_{f\in\funcs} \sum_{d\in\devs_c} \tau_{i,f,d} \geq 1\}$ contains all the requests that involve nodes belonging to cluster $c$ only;

\noindent
$\bullet$ \textit{Step III}: We solve Problem~\eqref{eq:problem} via B\&B considering only requests in $\reqs_c$ and nodes in $\devs_c$. The solution is a tuple $(\mathbf{x}_c,\mathbf{y}_c,\mathbf{z}_c)$ specifying which models are instantiated and where ($\mathbf{x}_c$), which requests are satisfied in cluster $c$ ($\mathbf{y}_c$) and what instances of the models are instantiated on each node of $\devs_c$ ($\mathbf{z}_c$).


\noindent
\textbf{Remark.}~This branching procedure might compute solutions with \textit{partially satisfied requests}. These are requests that are accommodated on a subset of clusters only, which violates Constraint \ref{eq:con1}. However, as we will show in Section \ref{sec:numerical}, this procedure is scalable as each subtree $\devs_c$ involves a limited number of nodes only, and we can solve each lower-dimensional instance of Problem \eqref{eq:problem} in parallel and in less than 0.1~s.


\section{Numerical Evaluation} \label{sec:numerical}


\begin{table}[t]
\renewcommand{\tabcolsep}{2.5pt}
\centering
\scriptsize
\caption{\scriptsize Controllable nodes.\vspace{-5pt}}
\label{tab:position}
\begin{tabular}{|l|c|c|c|c|c|}
\cline{1-6}
\backslashbox{Case}{Requested Nodes} & \nonrt RIC  & \nearrt RIC  & CU  & DU  & RU \\
\hline
\multicolumn{1}{|l|}{All nodes (ALL)} & \cmark       & \cmark        & \cmark & \cmark & \cmark \\ \hline
\multicolumn{1}{|l|}{Edge and RAN (ER)} & \xmark       & \cmark        & \cmark & \cmark & \cmark \\ \hline
\multicolumn{1}{|l|}{RAN only (RO)} & \xmark       & \xmark           & \cmark & \cmark & \cmark \\ \hline
\end{tabular}%
%
\vspace{5pt}
%
\caption{\scriptsize Request timescale cases and probabilities.\vspace{-5pt}}
\label{tab:time-scale}
\begin{tabular}{|l|c|c|c|}
\cline{1-4}
                         \backslashbox{Case}{Time scale~~~} & TTI-level - $\leq$ 0.01s & Sub-second - $\leq$ 1s & Long  - $>$ 1s \\\hline
\multicolumn{1}{|c|}{Delay-Tolerant (DT)} & 0.2       & 0.2         & 0.6                   \\ \hline
\multicolumn{1}{|c|}{Low Latency (LL)} & 0.2       & 0.6         & 0.2                   \\ \hline
\multicolumn{1}{|c|}{Ultra-Low Latency (ULL)} & 0.6       & 0.4         & 0                     \\ \hline
\end{tabular}

\end{table}

To evaluate the performance of \name in large-scale scenarios, we have developed a simulation tool in MATLAB that uses CPLEX to execute optimization routines.
For each simulation, \glspl{no} submit $R\!=\!20$ randomly generated requests, each specifying multiple sets of functionalities and nodes, as well as the desired timescale. Unless otherwise stated, we consider a single-domain deployment with 1 \nonrt \gls{ric}, 4 \nearrt \glspl{ric}, 10 \glspl{cu}, 30 \glspl{du} and 90 \glspl{ru}. For each simulation, the number of network nodes is fixed, but the tree structure of the infrastructure is randomly generated.
We consider the three cases shown in Table~\ref{tab:position}, where we limit the type of nodes that can be included in each request.
%
Similarly, we also consider the three cases in Table~\ref{tab:time-scale}. For each case, we specify the probability that the latency requirement $\delta_{i,f,d}$ for each tuple $(i,f,d)$ is associated to a specific timescale.
The combination of these 6 cases covers relevant Open RAN applications.

The ML/AI Catalog consists of $M=13$ models that provide $F=7$ different functionalities.
Ten models use metrics from the \gls{ran} (e.g., throughput and buffer measurements) as input, while the remaining three models are fed with IQ samples from \glspl{ru}.
The input size $s_{\tin_m}$ is set to 100 and 1000 bytes for the metrics and IQ samples, respectively.
For the sake of illustration, we assume that $\beta_{m,f,d} = \sigma_{m,f}$, $\pi_{i,f,d} = \tau_{i,f,d}$ and $C_{m,d}=3$ for all $m\in\mods$, $i\in\funcs$, $f\in\funcs$ and $d\in\devs$. The execution time of each model is equal across all models and nodes and set to $T^{exec}_{m,d}=1$~ms. The available bandwidth  $b_{d,d'}$ is 100~Gbps between \nonrt \gls{ric} and \nearrt \gls{ric}, 50~Gbps between \nearrt \glspl{ric} and \glspl{cu}, 25~Gbps between \glspl{cu} and \glspl{du}, and 20~Gbps between \glspl{du} and \glspl{ru}, while the propagation delay $T_{d,d'}$ is set to $[10, 10, 5, 1]$~ms, respectively.
The resources $\rho_d$ available at each node are represented by the number of available CPU cores, and we assume that each model requests one core only, i.e., $\rho_m = 1$. The number of cores available at \nonrt \glspl{ric}, \nearrt \glspl{ric}, \glspl{cu}, \glspl{du} and \glspl{ru} are 128, 8, 4, 2, and 1, respectively. Results presented in this section are averaged over 100 independent simulation runs.

\begin{figure}[t]
\begin{centering}
    \includegraphics[width=0.9\columnwidth]{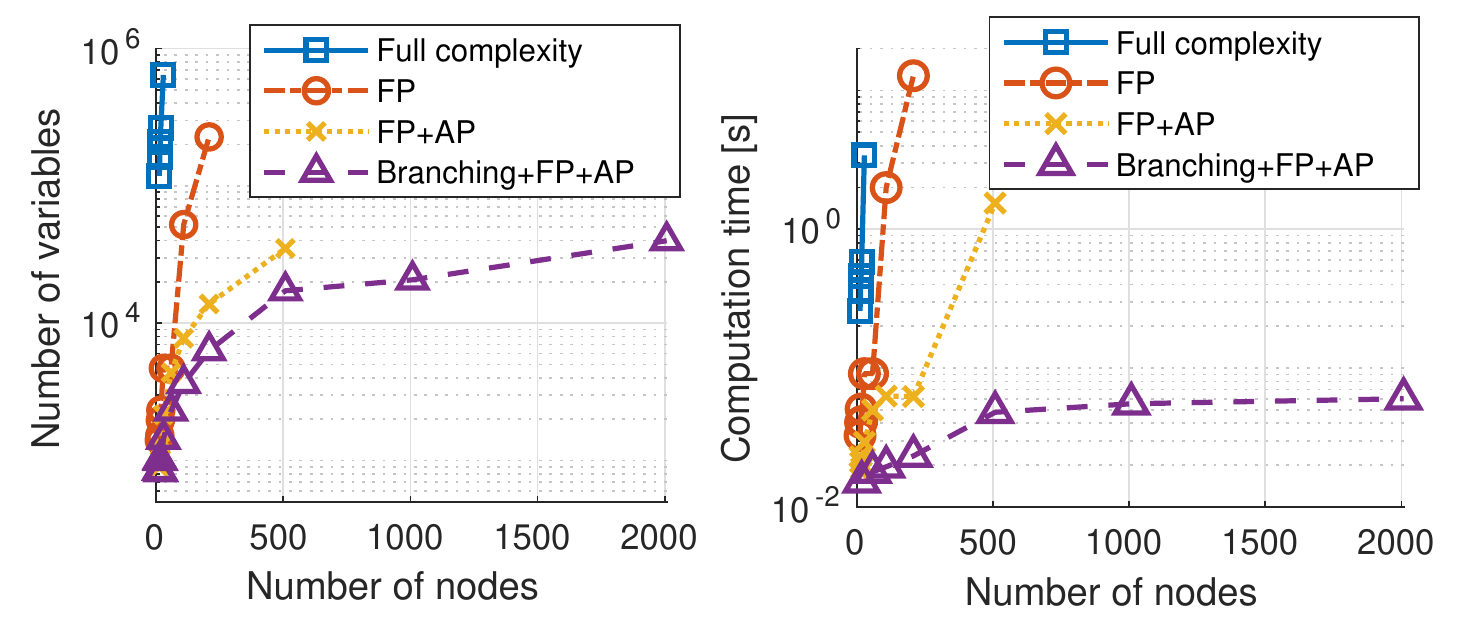}
    \vspace{-0.4cm}
    \caption{Number of variables and computation time for different network size.}
    \label{fig:scalability}
\end{centering}
\end{figure}

\noindent
$\bullet$~\textit{Computational complexity.}
%
Fig.~\ref{fig:scalability} shows the number of optimization variables and computation time of our algorithms with varying network size. At each simulation run, we consider a single \nonrt \gls{ric} and a randomly generated tree graph that matches the considered size. As expected, the number of variables and the complexity increase with larger networks. This can be mitigated by using our FP and AP pre-processing algorithms, which reduce the number of optimization variables while ensuring the optimality of the computed solution.
Their combination allows computation of optimal solutions in 0.1~s and 2~s for networks with 200 and 500 nodes, respectively.
Fig.~\ref{fig:scalability} also shows the benefits of branching the optimization problem into sub-problems of smaller size (Section \ref{sec:branching}).
\begin{figure}[b]
\begin{centering}
    \includegraphics[width=0.9\columnwidth]{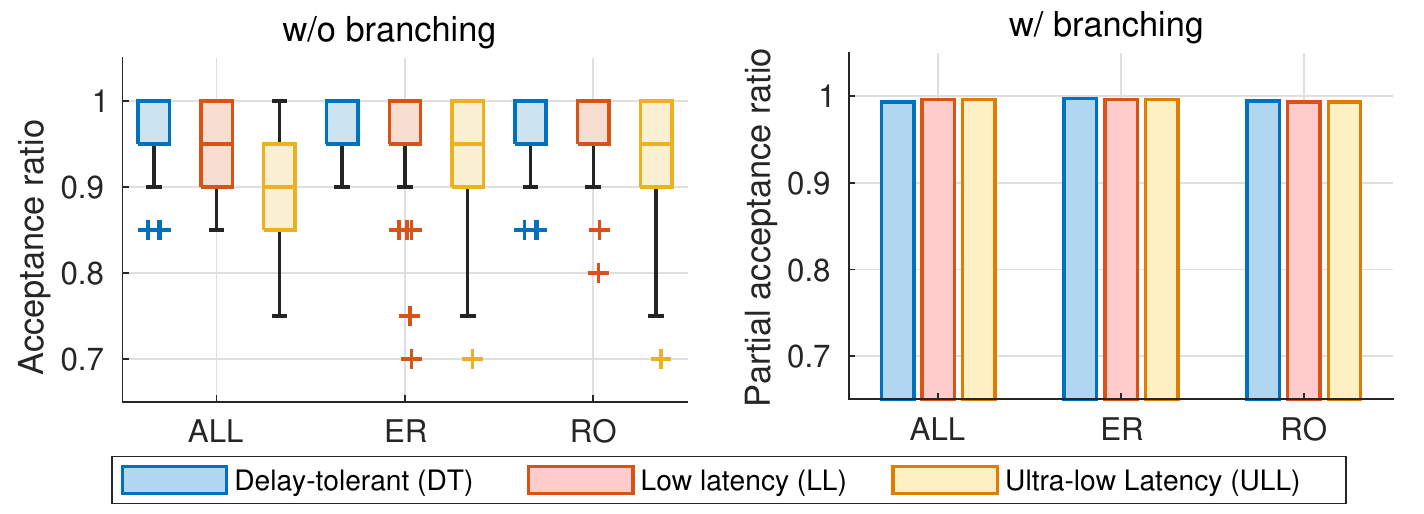}
    \vspace{-0.4cm}
    \caption{(Left) Ratio of accepted requests w/ model sharing but w/o branching; (Right) Ratio of partially accepted requests w/ model sharing and branching.}
    \label{fig:acceptance}
\end{centering}
\end{figure}
Although the branching procedure might produce partially satisfied requests, it results in a computation time lower than 0.1~s even for instances with 2000 nodes, providing a fast and scalable solution for large-scale applications.

\noindent
$\bullet$~\textit{Acceptance ratio.} Fig.~\ref{fig:acceptance} (left) shows the acceptance ratio for different cases and algorithms.
The number of accepted requests decreases when moving from loose timing requirements (i.e., Delay-tolerant (DT)), to tighter ones (i.e., Low Latency (LL) and Ultra-Low Latency (ULL)).
For example, while 95\% of requests are satisfied on average for the DT configuration, we observe ULL instances in which only 70\% of requests are accepted. Indeed, \gls{tti}-level services may only be possible at the \glspl{du}/\glspl{ru} which, however, have limited resources and cannot support the execution of many concurrent \oran applications.
In Fig. \ref{fig:acceptance} (right), we show the probability that a request is partially accepted when considering the branching algorithm.
%
%
Specifically, it shows that branching results in $\approx$99\% of requests being partially satisfied on one subtree or more. This means that in the case where not enough resources are available to accept the entirety of the request, \name can satisfy portions of it. Thus, requests that would be otherwise rejected can be at least partially accommodated.

\begin{figure}[t]
\begin{centering}
    \includegraphics[width=0.9\columnwidth]{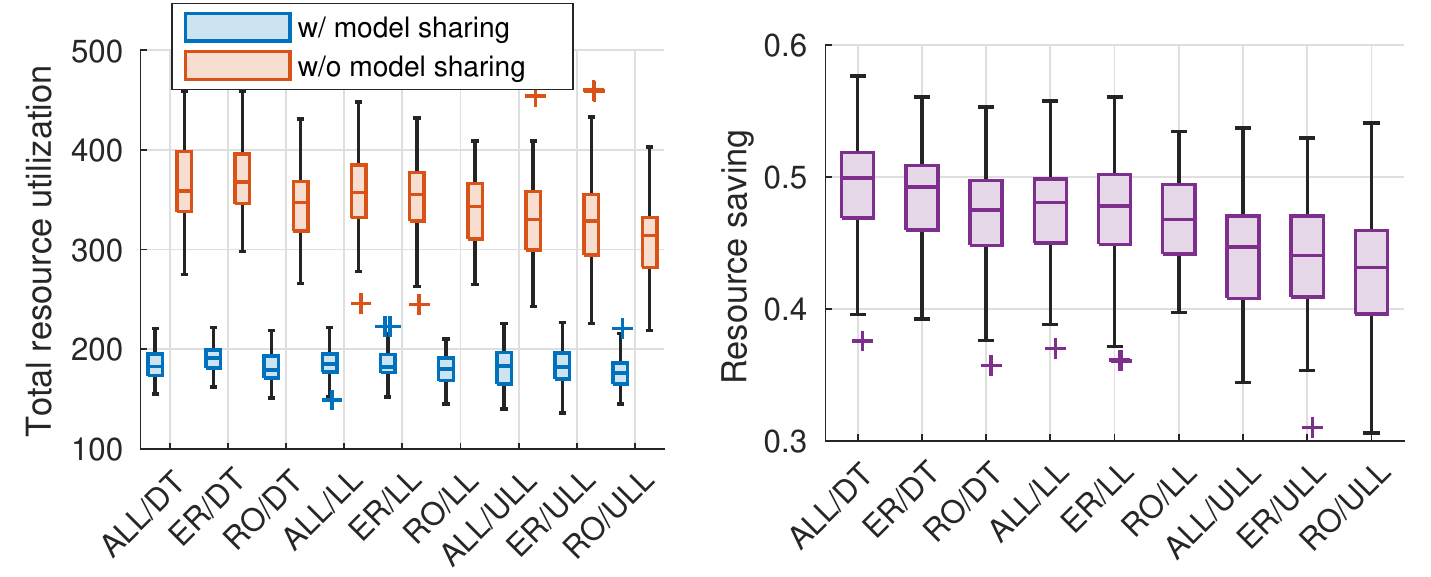}
    \vspace{-0.4cm}
    \caption{Resource utilization and saving with and without model sharing.}
    \label{fig:sharing_resources}
\end{centering}
\end{figure}

\begin{figure}[b]
\centering
\begin{minipage}[t]{0.35\columnwidth}
\begin{centering}
    \includegraphics[width=\columnwidth]{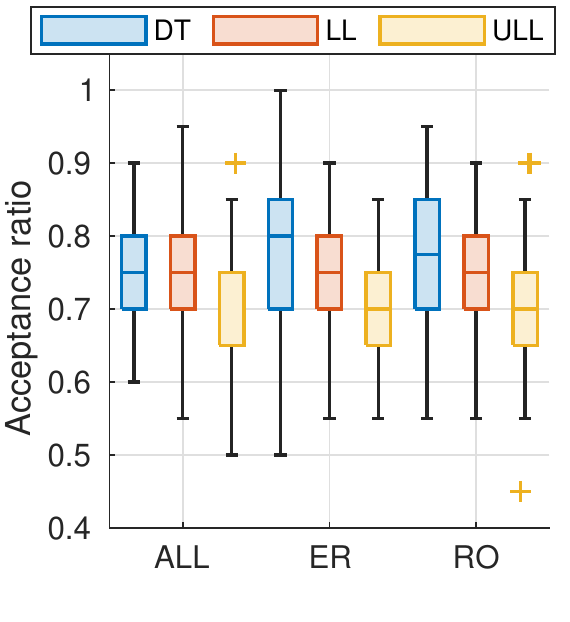}
    \vspace{-0.7cm}
    \caption{\label{fig:acceptance_no_sharing}Percentage of accepted requests w/o model sharing for different cases.}
\end{centering}
\end{minipage}
\hspace{2pt}
\begin{minipage}[t]{0.62\columnwidth}
\begin{centering}
    \includegraphics[width=\columnwidth]{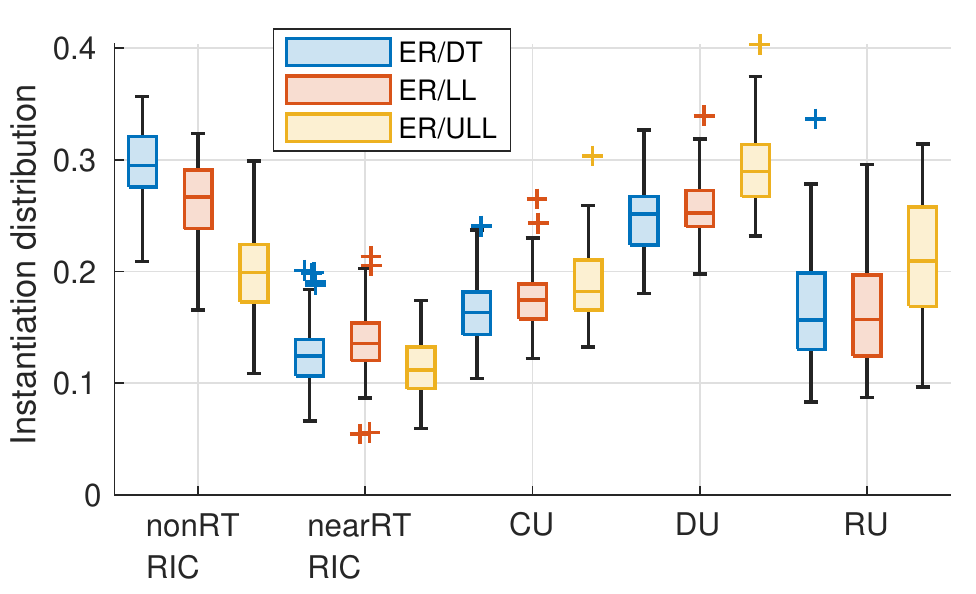}
    \vspace{-0.7cm}
    \caption{\label{fig:instantiation}Distribution of model instantiation for different cases.}
\end{centering}
\end{minipage}
\end{figure}

\noindent
$\bullet$~\textit{Advantages of model sharing.}
Fig.~\ref{fig:sharing_resources} shows the resource utilization with and without model sharing (left) and the corresponding resource utilization saving (right).
As expected, model sharing always results in lower resource utilization and uses 2$\times$ less resources than the case without model sharing.
%
Fig.~\ref{fig:acceptance_no_sharing} shows the acceptance ratio when model sharing is disabled, and by comparing it with Fig.~\ref{fig:acceptance} (left)---where model sharing is enabled---we notice that model sharing also increases the acceptance ratio.
Specifically, model sharing accommodates at least 90\% of requests in all cases, while this number drops to $\approx$70\% when model sharing is disabled.
%

To better understand how \name orchestrates intelligence, Fig.~\ref{fig:instantiation} shows the distribution of models across the different network nodes for the ER case (see Table~\ref{tab:position}) with different timing constraints. Requests with loose timing requirements (DT) result in $\approx$45\% of models being allocated in the \glspl{ric}. Instead, stringent timing constraints (LL and ULL) result in $\approx$70\% of models being instantiated at \glspl{cu}, \glspl{du}, and \glspl{ru}.

\section{Prototype and Experimental Evaluation} \label{sec:experiments}

To demonstrate the effectiveness of \name, we leveraged ColO-RAN~\cite{polese2021coloran}, an \oran-compliant large-scale experimental platform developed on top of the Colosseum wireless network emulator~\cite{bonati2021colosseum}.
Colosseum includes 128 computing servers (i.e., \glspl{srn}), each controlling a USRP~X310 \gls{sdr}, and a \gls{mchem} emulating wireless channels between the \glspl{srn}
%
to reproduce realistic and time-varying wireless characteristics (e.g., path-loss, multi-path) under different deployments (e.g., urban, rural, etc.)~\cite{bonati2021colosseum}.
We leverage the publicly available tool SCOPE~\cite{bonati2021scope} to instantiate a softwarized cellular network with 7 base stations and 42 \glspl{ue} (6~\glspl{ue} per base station) on the Colosseum city-scale downtown Rome scenario, and to interface the base stations with the \oran \nearrt \gls{ric} through the E2 interface. SCOPE, which is based on srsRAN~\cite{gomez2016srslte}, implements open \glspl{api} to reconfigure the base station parameters (e.g., slicing resources, scheduling policies, etc.) from \oran applications through closed-control loops, and to automatically generate datasets from \gls{ran} statistics (e.g., throughput, buffer size, etc.).
Users are deployed randomly and generate traffic belonging to 3 different network slices configured as follows: (i)~slice~0 is allocated an \gls{embb} service, in which each \gls{ue} requests 4~Mbps constant bitrate traffic; (ii)~slice~1 a \gls{mtc} service, in which each \gls{ue} requests Poisson-distributed traffic with an average rate of 45~kbps, and (iii)~slice~2 to a \gls{urllc} service, in which each \gls{ue} requests Poisson-distributed traffic with an average rate of 90~kbps. We assume 2~\glspl{ue} per slice, whose traffic is handled by the base stations, which use a 10~MHz channel bandwidth with 50 \gls{prb}.

\begin{table}[t]
\renewcommand{\tabcolsep}{2.5pt}
\caption{\scriptsize \gls{drl} agents in the ML/AI Catalog}
\vspace{-0.2cm}
\scriptsize
\label{tab:drl}
\begin{tabular}{l|c|c|c|c}
\cline{2-4}
\multicolumn{1}{c|}{} & \multicolumn{3}{c|}{Reward} &                              \\ \cline{2-5}
                      & Slice 0 & Slice 1 & Slice 2 & \multicolumn{1}{c|}{Actions} \\ \hline
\multicolumn{1}{|l|}{M3} & max(Throughput) & max(TX pkts) & max(PRB ratio)   & \multicolumn{1}{c|}{Scheduling}                     \\ \hline
\multicolumn{1}{|l|}{M4} & max(Throughput) & max(TX pkts) & min(Buffer size) & \multicolumn{1}{c|}{Scheduling, \gls{ran} slicing } \\ \hline
\end{tabular}
\end{table}

\begin{figure}[t]
\begin{centering}
    \includegraphics[width=0.9\columnwidth]{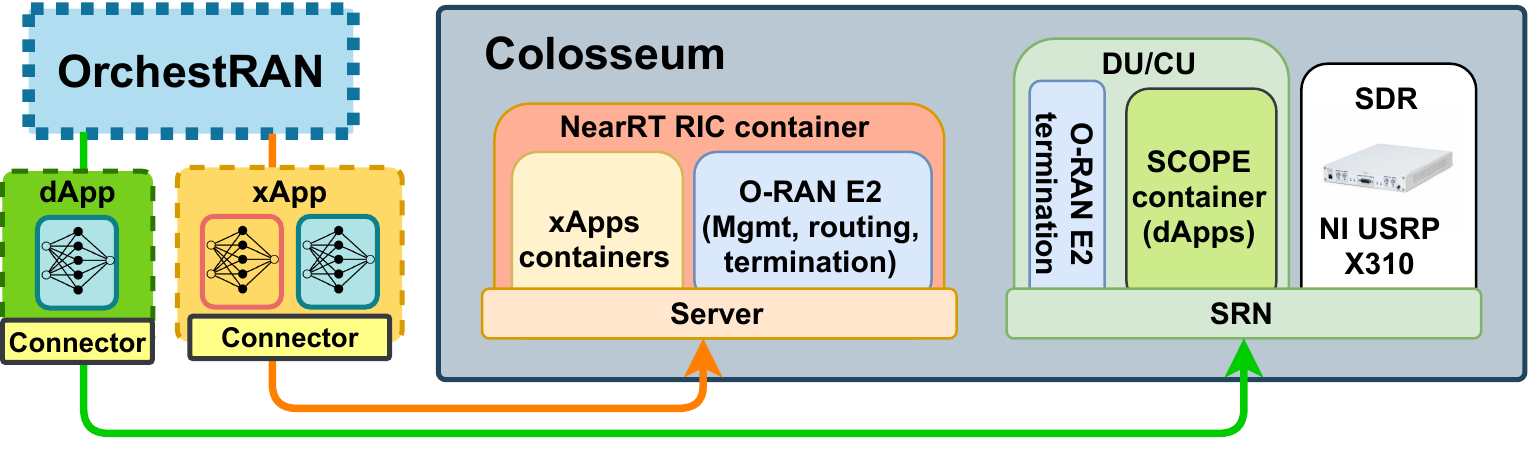}
    \vspace{-0.4cm}
    \caption{\name prototype architecture on Colosseum and integration with \oran and SCOPE~\cite{bonati2021scope} components.}
    \label{fig:prototype}
\end{centering}
\end{figure}

The high-level architecture of the \name prototype on Colosseum is shown in Fig.~\ref{fig:prototype}. \name runs in an \gls{lxc} embedding the components of Fig~\ref{fig:system}. For each experiment, we randomly generate a new set of control requests every 4~minutes. The Orchestration Engine computes the optimal orchestration policy and embeds the models within \oran applications that are dispatched to the nodes where they are executed.
We consider the case where models can run at the \nearrt \gls{ric} (as xApps) or at the \gls{du} (as dApps via SCOPE).

We used SCOPE to generate datasets on Colosseum and train 4 ML models that constitute our ML/AI Catalog.
Models $M1$ and $M2$ have been trained to forecast throughput and transmission buffer size.\footnote{Due to space limitations, and since the goal of this paper is focused on how to orchestrate pre-trained models to accomplish \glspl{no} goals, details on the training procedures are omitted.} Models $M3$ and $M4$ control the parameters of the network to maximize different rewards through \gls{ppo}-based \gls{drl} agents (see Table~\ref{tab:drl}).
%
Specifically, $M3$ consists of three \gls{drl} agents, each making decisions on the scheduling policies of one slice only. The three agents aim at maximizing the throughput of slice~0, the number of transmitted packets of slice~1, and the ratio between the allocated and requested \glspl{prb} (i.e., the \textit{\gls{prb} ratio} which takes values in $[0,1]$) of slice~2, respectively.
Model $M4$, instead, consists of a single \gls{drl} agent controlling the scheduling and \gls{ran} slicing policies (i.e., how many \glspl{prb} are assigned to each slice) to \textit{jointly} maximize the throughput of slice~0 and the number of transmitted packets of slice~1, and to minimize the buffer size of slice~2. Each model requires one CPU core, and we consider three configurations: (i)~``\gls{ric} only'', in which models can be executed via xApps at the \nearrt \gls{ric} only; (ii)~``\gls{ric} + lightweight \gls{du}'', in which \glspl{du} have 2 cores each to execute up to two dApps concurrently; and (iii)~``\gls{ric} + powerful \gls{du}'', in which \glspl{du} are equipped with 8 cores. In all cases, the \nearrt \gls{ric} has access to 50 cores.
Overall, we ran more than 95 hours of experiments on Colosseum.

\begin{figure}[t]
\begin{centering}
    \includegraphics[width=0.9\columnwidth]{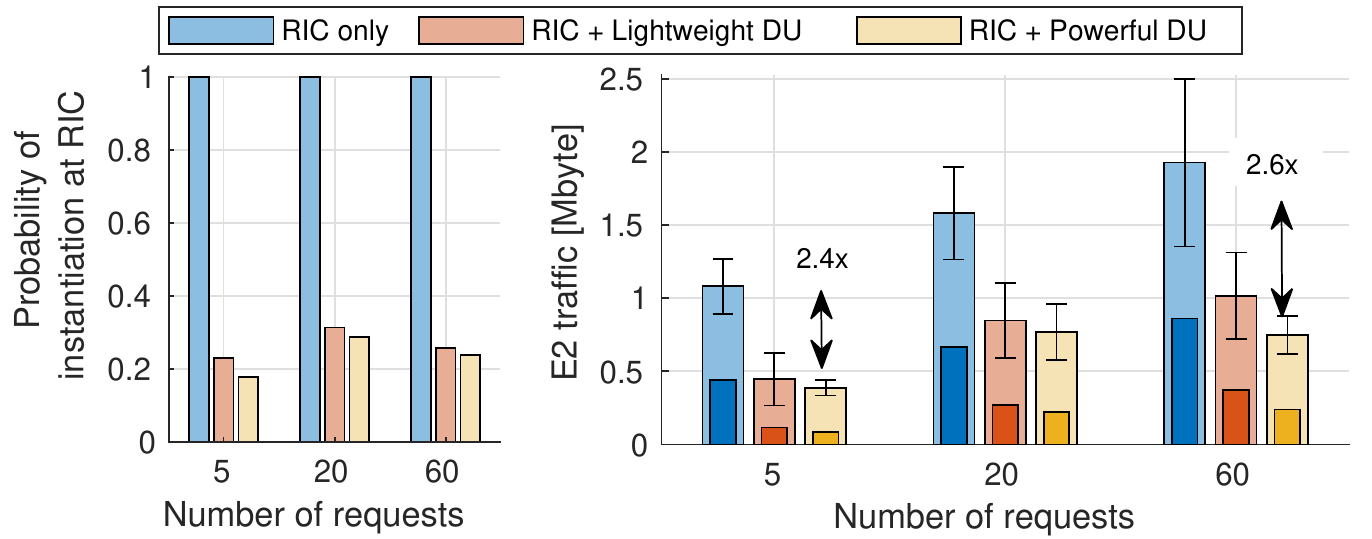}
    \vspace{-0.3cm}
    \caption{(Left) Probability of instantiating \oran applications at the \nearrt \gls{ric}; (Right) Traffic over \oran E2 interface for different configurations. Dark bars represent traffic related to payload only.}
    \label{fig:e2}
\end{centering}
\end{figure}

\noindent
$\bullet$~\textit{Experimental results.}~Fig.~\ref{fig:e2} (left) shows the probability that models are executed at the \nearrt \gls{ric} for different configurations and number of requests.
As expected, in the ``\gls{ric} only" case, all models execute as xApps at the \nearrt \gls{ric}, while both ``RIC + lightweight \gls{du}" and ``RIC + powerful \gls{du}" cases result in $\approx$25\% of models executing at the \gls{ric}. The remaining 75\% of the models are executed as dApps at the \glspl{du}.
%
Fig. \ref{fig:e2} (right) shows the traffic in Mbyte over the E2 interface between the \nearrt \gls{ric} and the \glspl{du} for the different configurations.
This includes messages to set up the initial subscription between the \nearrt \gls{ric} and the \glspl{du}, messages to report metrics from the \glspl{du} to the \gls{ric} (e.g., throughput, buffer size), and control messages from the \gls{ric} to the \glspl{du} (e.g., to update scheduling and \gls{ran} slicing policies).
%
%
Results clearly show that $\approx$40\% of the E2 traffic transports payload information (dark bars), while the remaining 60\% consists of overhead data. Although the initial subscription messages exchanged between the \nearrt \gls{ric} and the \glspl{du} are sent in all considered cases, running models as dApps at the \glspl{du} still results in up to 2.6$\times$ less E2 traffic if compared to the ``\gls{ric} only" case.
%

\begin{figure}[t]
\begin{centering}
    \includegraphics[width=0.85\columnwidth]{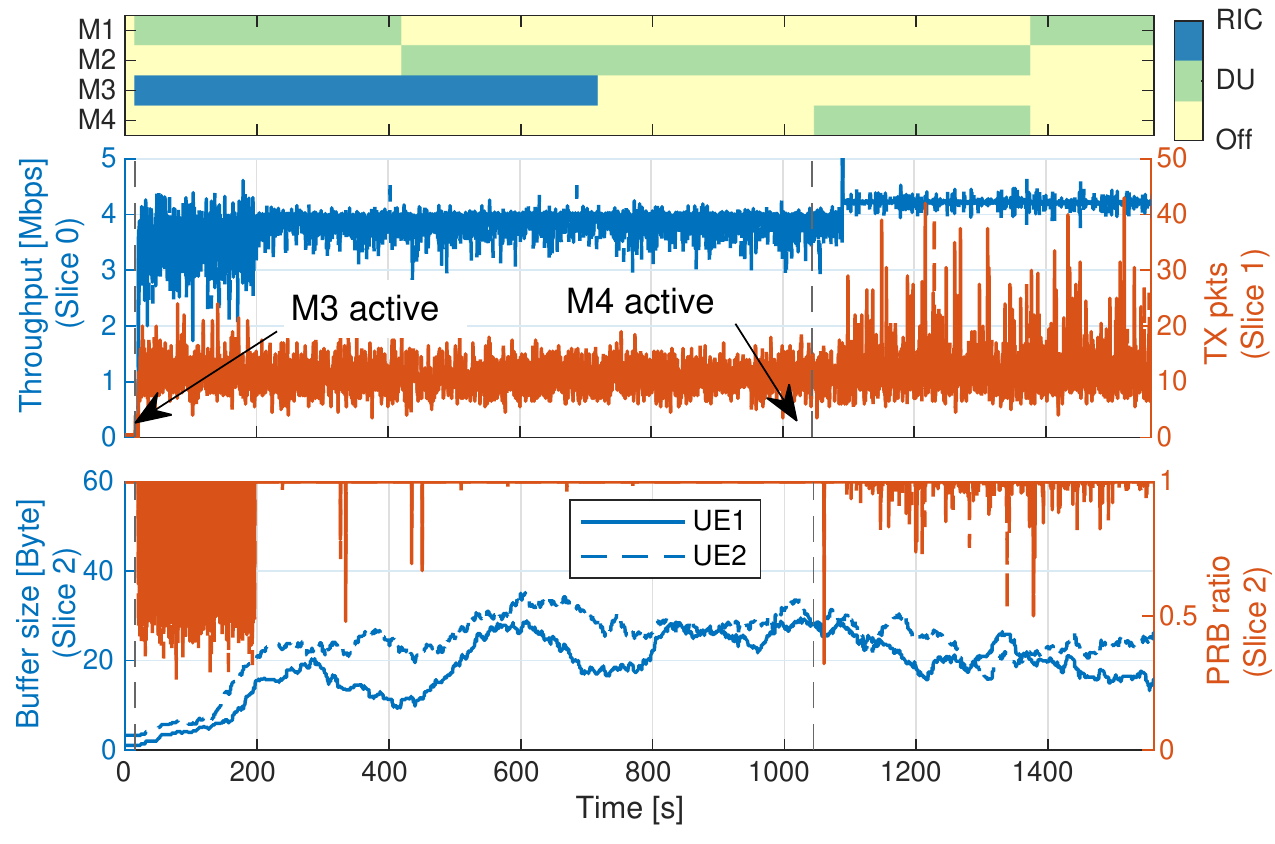}
    \vspace{-0.3cm}
    \caption{(Top) Dynamic activation of \oran applications at \nearrt \gls{ric} and \gls{du} 7; (Center and bottom) Performance comparison for different deployments of \oran applications and network slices. Solid lines and dashed lines refer to traffic for $\mathrm{UE}_{1,i}$ and $\mathrm{UE}_{2,i}$ of Slice $i$.}
    \label{fig:container}
\end{centering}
\end{figure}

Finally, we showcase the impact of the real-time execution of \name on the network performance.
We focus on \gls{du}~7, and in Fig.~\ref{fig:container} (top) we show the location and time instant at which \name instantiates the four models on the \nearrt \gls{ric} and on \gls{du}~7 for a single experiment.
%
The impact on the network performance of the different orchestration policies is shown in Fig.~\ref{fig:container} (center and bottom). Since $M1$ and $M2$ perform forecasting tasks only, the figure only reports the evolution of the metrics used to reward the \gls{drl} agents $M3$ and $M4$ (see Table~\ref{tab:drl}) for different slices.
%
We notice that \name allows the seamless instantiation of dApps and xApps, controlling the same \gls{du} without causing any service interruptions.
Moreover, although $M3$ and $M4$ share the same reward for slices~0 and~1, $M4$ can also make decisions on the network slicing policies. Thus, it provides a higher throughput for slice~0 ($\approx$10\% higher than $M3$), and a higher number of transmitted packets for slice~1 ($\approx$2$\times$ higher than $M3$) (Fig. \ref{fig:container} (center)).
Similarly, in the case of slice~2, $M3$ aims at maximizing the \gls{prb} ratio, while $M4$ at minimizing the size of the transmission buffer, which results in $M3$ and $M4$ computing different control policies for slice~2. As shown Fig.~\ref{fig:container} (bottom), although $M3$ converges to a stable control policy that results in a \gls{prb} ratio $\approx$1, its buffer size is higher than that of $M4$. Conversely, the buffer size of slice~2 decreases once $M4$ is instantiated with a decrease in the \gls{prb} ratio.


\section{Conclusions} \label{sec:conclusions}
In this paper, we presented \name, a novel network intelligence orchestration framework for Open RAN systems. \name is based upon \oran specifications and leverages the \gls{ric} xApps and rApps and \oran open interfaces to provide \glspl{no} with an automated orchestration tool for deploying data-driven inference and control solutions with diverse timing requirements. \name has been equipped with orchestration algorithms with different optimality/complexity trade-offs to support \nonrt, \nearrt and \rt applications.
We assessed \name performance and presented an \oran-compliant prototype by instantiating a cellular network with 7~base stations and 42~\glspl{ue} on the Colosseum network emulator. Our experimental results demonstrate that \name achieves seamless instantiation of \oran applications at different network nodes and time scales, and reduces the message overhead over the \oran E2 interface by up to 2.6$\times$ when instantiating intelligence at the edge of the network.

\balance
\footnotesize
\bibliographystyle{IEEEtran}
\bibliography{bibliography}

\end{document}